# An Introduction to Programming for Bioscientists: A Python-based Primer

Berk Ekmekci[1,☯], Charles McAnany[1,☯], Cameron Mura[1, *]

**1 Department of Chemistry, University of Virginia, Charlottesville, VA 22904-4319 USA**

☯These authors contributed equally to this work.
*cmura@muralab.org

## Abstract

Computing has revolutionized the biological sciences over the past several decades, such that virtually all contemporary research in molecular biology, biochemistry, and other biosciences utilizes computer programs. The computational advances have come on many fronts, spurred by fundamental developments in hardware, software, and algorithms. These advances have influenced, and even engendered, a phenomenal array of bioscience fields, including molecular evolution and bioinformatics; genome–, proteome–, transcriptome– and metabolome–wide experimental studies; structural genomics; and atomistic simulations of cellular-scale molecular assemblies as large as ribosomes and intact viruses. In short, much of post-genomic biology is increasingly becoming a form of computational biology. The ability to design and write computer programs is among the most indispensable skills that a modern researcher can cultivate. Python has become a popular programming language in the biosciences, largely because (i) its straightforward semantics and clean syntax make it a readily accessible first language; (ii) it is expressive and well-suited to object-oriented programming, as well as other modern paradigms; and (iii) the many available libraries and third-party toolkits extend the functionality of the core language into virtually every biological domain (sequence and structure analyses, phylogenomics, workflow management systems, etc.). This primer offers a basic introduction to coding, via Python, and it includes concrete examples and exercises to illustrate the language's usage and capabilities; the main text culminates with a final project in structural bioinformatics. A suite of Supplemental Chapters is also provided. Starting with basic concepts, such as that of a 'variable', the Chapters methodically advance the reader to the point of writing a graphical user interface to compute the Hamming distance between two DNA sequences.

## Author Summary

Contemporary biology has largely become computational biology, whether it involves applying physical principles to simulate the motion of each atom in a piece of DNA, or using machine learning algorithms to integrate and mine 'omics' data across whole cells (or even entire ecosystems). The ability to design algorithms and program computers, even at a novice level, may be the most indispensable skill that a modern researcher can cultivate. As with human languages, computational fluency is developed actively, not passively. This self-contained text, structured as a hybrid primer/tutorial, introduces any biologist—from college freshman to established senior scientist—to basic computing principles (control-flow, recursion, regular expressions, etc.) and the practicalities of programming and software design.

We use the Python language because it now pervades virtually every domain of the biosciences, from sequence-based bioinformatics and molecular evolution to phylogenomics, systems biology, structural biology, and beyond. To introduce both coding (in general) and Python (in particular), we guide the reader via concrete examples and exercises. We also supply, as Supplemental Chapters, a few thousand lines of heavily-annotated, freely distributed source code for personal study.

# Introduction

## Motivation: Big Data & Biology

Datasets of unprecedented volume and heterogeneity are becoming the norm in science, and particularly in the biosciences. High-throughput experimental methodologies in genomics [1], proteomics [2], transcriptomics [3], metabolomics [4], and other 'omics' [5–7] routinely yield vast stores of data on a system-wide scale. Growth in the quantity of data has been matched by an increase in heterogeneity: There is now great variability in the *types* of relevant data, including nucleic acid and protein sequences from large-scale sequencing projects, proteomic data and molecular interaction maps from microarray and chip experiments on entire organisms (and even ecosystems [8–10]), three-dimensional (3D) coordinate data from international structural genomics initiatives, petabytes of trajectory data from large-scale biomolecular simulations, and so on. In each of these areas, volumes of raw data are being generated at rates that dwarf the scale and exceed the scope of conventional data-processing and data-mining approaches.

The intense data-analysis needs of modern research projects feature at least three facets: Data *production*, *reduction/processing*, and *integration*. Data production is largely driven by engineering and technological advances, such as commodity equipment for next-gen DNA sequencing [11–13] and robotics for structural genomics [14, 15]. Data reduction requires efficient computational processing approaches, and data integration demands robust tools that can flexibly represent data (*abstractions*) so as to enable the detection of correlations and interdependencies (via, e.g., machine learning [16]). These facets are closely coupled: The rate at which raw data is now produced, e.g. in computing molecular dynamics (MD) trajectories [17], dictates the data storage, processing, and analysis needs. As a concrete example, the latest generation of highly-scalable, parallel MD codes can generate data more rapidly than they can be transferred, via typical computer network backbones, to local workstations for processing. Such demands have spurred the development of tools for 'on-the-fly' trajectory analysis (e.g., [18, 19]) as well as generic software toolkits for constructing parallel and distributed data-processing pipelines (e.g., [20] and S2 Text, §2). To appreciate the scale of the problem, note that calculation of all-atom MD trajectories over biologically-relevant timescales easily leads into petabyte-scale computing. Consider, for instance, a biomolecular simulation system of modest size, such as a 100-residue globular protein embedded in explicit water (corresponding to $\approx 10^5$ particles), and with typical simulation parameters (32-bit precision, atomic coordinates written to disk, in binary format, for every ps of simulation time, etc.). Extending such a simulation to $10\,\mu s$ duration—which may be at the low end of what is deemed biologically relevant for the system—would give an approximately 12-terabyte trajectory ($\approx 10^5$ particles $\times$ 3 coordinates/particle/frame $\times\, 10^7$ frames $\times$ 4 bytes/coordinate $= 12\,\text{TB}$). To validate or otherwise follow-up predictions from a single trajectory, one might like to perform an additional suite of $>10$ such simulations, thus rapidly approaching the peta-scale.

Scenarios similar to the above example occur in other biological domains, too, at length-scales ranging from atomic to organismal. Atomistic MD simulations were mentioned above. At the molecular level of individual genes/proteins, an early step in characterizing a protein's function and evolution might be to use sequence analysis methods to compare the protein sequence to every other known sequence, of which there are tens of millions [21]. Any form of 3D structural analysis will almost certainly involve the Protein Data Bank (PDB; [22]), which currently holds over $10^5$ entries. At the cellular level, proteomics, transcriptomics, and various other 'omics' areas (mentioned above) have been inextricably linked to high-throughput, big-data science since the inception of each of those fields. In genomics, the early bottleneck—DNA sequencing and raw data collection—was eventually supplanted by the problem of processing raw sequence data into derived (secondary) formats, from which point meaningful conclusions can be gleaned [23]. Enabled by the amount of data that can be rapidly generated, typical 'omics' questions have become more subtle. For instance, simply assessing sequence similarity and conducting functional annotation of the open reading frames (ORFs) in a newly sequenced genome is no longer the end-goal; rather, one might now seek to derive networks of biomolecular functions from sparse, multi-dimensional datasets [24]. At the level of tissue systems, the modeling and simulation of inter-neuronal connections has developed into a new field of 'connectomics' [25, 26]. Finally, at the organismal and clinical level, the promise of personalized therapeutics hinges on the ability to analyze large, heterogeneous collections of data (e.g., [27]). As illustrated by these examples, all bioscientists would benefit from a basic understanding of the computational tools that are used daily to collect, process, represent, statistically manipulate, and otherwise analyze data. In every data-driven project, the overriding goal is to transform raw data into new biological principles and knowledge.

## A New Kind of Scientist

Generating knowledge from large datasets is now recognized as a central challenge in science [28]. To succeed, each type of aforementioned data-analysis task hinges upon three things: greater computing power, improved computational methods, and computationally fluent scientists. Computing power is only marginally an issue: It lies outside the scope of most biological research projects, and the problem is often addressed by money and the acquisition of new hardware. In contrast, computational methods—improved algorithms, and the software engineering to implement the algorithms in high-quality codebases—are perpetual goals. To address the challenges, a new era of scientific training is required [29–32]. There is a dire need for biologists who can collect, structure, process/reduce, and analyze (both numerically and visually) large-scale datasets. The problems are more fundamental than, say, simply converting data files from one format to another ('data-wrangling'). Fortunately, the basics of the necessary computational techniques can be learned quickly. Two key pillars of computational fluency are (i) a working knowledge of some programming language and (ii) comprehension of core computer science principles (data structures, sort methods, etc.). All programming projects build upon the same set of basic principles, so a seemingly crude grasp of programming essentials will often suffice for one to understand the workings of very complex code; one can develop familiarity with more advanced topics (graph algorithms, computational geometry, numerical methods, etc.) as the need arises for particular research questions. Ideally, computational skills will begin to be developed during early scientific training. Recent educational studies have exposed the gap in life sciences and computer science knowledge among young scientists, and interdisciplinary education appears to be effective in helping bridge the gap [33, 34].

## Programming as the Way Forward

For many of the questions that arise in research, software tools have been designed. Some of these tools follow the Unix tradition to "Make each program do one thing well" [35], while other programs have evolved into colossal applications that provide numerous sophisticated features, at the cost of accessibility and reliability. A small software tool that is designed to perform a simple task will, at some point, lack a feature that is necessary to analyze a particular type of dataset. A large program may provide the missing feature, but the program may be so complex that the user cannot readily master it, and the codebase may have become so unwieldy that it cannot be adapted to new projects without weeks of study. Guy Steele, a highly-regarded computer scientist, noted this principle in a lecture on programming language design [36]:

> *"I should not design a small language, and I should not design a large one. I need to design a language that can grow. I need to plan ways in which it might grow—but I need, too, to leave some choices so that other persons can make those choices at a later time."*

Programming languages provide just such a tool. Instead of supplying every conceivable feature, languages provide a small set of well-designed features and powerful tools to compose these features in new ways, using logical principles. Programming allows one to control every aspect of data analysis, and libraries provide commonly-used functionality and pre-made tools that the scientist can use for most tasks. A good library provides a simple interface for the user to perform routine tasks, but also allows the user to tweak and customize the behavior in any way desired (such code is said to be *extensible*). The ability to compose programs into other programs is particularly valuable to the scientist. One program may be written to perform a particular statistical analysis, and another program may read in a data file from an experiment and then use the first program to perform the analysis. A third program might select certain datasets—each in its own file—and then call the second program for each chosen data file. In this way, the programs can serve as modules in a computational workflow.

On a related note, many software packages supply an application programming interface (API), which exposes some specific set of functionalities from the codebase without requiring the user/programmer to worry about the low-level implementation details. A well-written API enables users to combine already established codes in a modular fashion, thereby more efficiently creating customized new tools and pipelines for data processing and analysis.

A program that performs a useful task can (and, arguably, *should* [37]) be distributed to other scientists, who can then integrate it with their own code. Free software licenses facilitate this type of collaboration, and explicitly encourage individuals to enhance and share their programs [38]. This flexibility and ease of collaborating allow scientists to develop software relatively quickly, so they can spend more time integrating and mining, rather than simply processing, their data.

Data-processing workflows and pipelines that are designed for use with one particular program or software environment will eventually be incompatible with other software tools or workflow environments; such approaches are often described as being *brittle*. In contrast, algorithms and programming logic, together with robust and standards-compliant data-exchange formats, provide a completely universal solution that is portable between different tools. Simply stated, any problem that can be solved by a computer can be solved using any programming language [39, 40]. The more feature-rich or *high-level* the language, the more concisely can a data-processing task be expressed using that language (the language is said to be *expressive*). Many high-level languages (e.g., Python, Perl) are executed by an *interpreter*, which is a program that reads source code and does what the code says to do. Interpreted languages are not as numerically efficient as lower-level, compiled languages such as C or Fortran. The source code of a program in a *compiled* language must be converted to machine-specific instructions by a compiler, and those low-level machine code instructions (*binaries*) are executed directly by the hardware. Compiled code typically runs faster than interpreted code, but requires more work to program. High-level languages, such as Python or Perl, are often used to prototype ideas or to quickly combine modular tools (which may be written in a lower-level language) into 'scripts'; for this reason they are also known as *scripting languages*. Very large programs often provide a scripting language for the user to run their own programs: Microsoft Office has the VBA scripting language, PyMOL [41] provides a Python interpreter, VMD [42] uses a Tcl interpreter for many tasks, and Coot [43] uses the Scheme language to provide an API to the end-user. The deep integration of high-level languages into packages such as PyMOL and VMD enables one to extend the functionality of these programs via both scripting commands (e.g., see PyMOL examples in [44]) and the creation of semi-standalone plugins (e.g., see the VMD plugin at [45]). While these tools supply interfaces to different programming languages, the fundamental concepts of programming are preserved in each case: A script written for PyMOL can be transliterated to a VMD script, and a closure in a Coot script is roughly equivalent to a closure in a Python script (see Supplemental Chapter 13 in S1 Text). Because the logic underlying computer programming is universal, mastering one language will open the door to learning other languages with relative ease. As another major benefit, the algorithmic thinking involved in writing code to solve a problem will often lead to a deeper and more nuanced understanding of the scientific problem itself.

## Why Python? (And Which Python?)

Python is the programming language used in this text because of its clear syntax [40, 46], active developer community, free availability, extensive use in scientific communities such as bioinformatics, its role as a scripting language in major software suites, and the many freely available scientific libraries (e.g., BioPython [47]). Two of these characteristics are especially important for our purposes: (i) A clean syntax and straightforward semantics allow the student to focus on core programming concepts without the distraction of difficult syntactic forms, while (ii) the widespread adoption of Python has led to a vast base of scientific libraries and toolkits for more advanced programming projects [20, 48]. As noted in the S2 Text (§1), several languages other than Python have also seen widespread use in the biosciences; see, e.g., [46] for a comparative analysis of some of these languages. As described by Hinsen [49], Python's particularly rapid adoption in the sciences can be attributed to its powerful and versatile combination of features, including characteristics intrinsic to the language itself (e.g., expressiveness, a powerful object model) as well as extrinsic features (e.g., community libraries for numerical computing).

Two versions of Python are frequently encountered in scientific programming: Python 2 and Python 3. The differences between these are minor, and while this text uses Python 3 exclusively, most of the code we present will run under both versions of Python. Python 3 is being actively developed and new features are added regularly; Python 2 support continues mainly to serve existing ('legacy') codes. New projects should use Python 3.

## Role, and Organization, of This Text

This work, which has evolved from a modular "Programming for Bioscientists" tutorial series that has been offered at our institution, provides a self-contained, hands-on primer for general-purpose programming in the biosciences. Where possible, explanations are provided for key foundational concepts from computer science; more formal, and comprehensive, treatments can be found in several computer science texts [39, 40, 50] as well as bioinformatics titles, from both theoretical [16, 51] and more practical [52–55] perspectives. Also, this work complements other practical Python primers [56], guides to getting started in bioinformatics (e.g., [57, 58]), and more general educational resources for scientific programming [59].

Programming fundamentals, including variables, expressions, types, functions, control flow and recursion, are introduced in the first half of the text (§2). The next major section (§3) presents data structures for collections of items (lists, tuples, dictionaries) and more control flow (loops). Classes, methods, and other basics of object-oriented programming (OOP) are described in §4. File management and input/output (I/O) is covered in §5, and another practical (and fundamental) topic associated with data-processing—regular expressions for string parsing—is covered in §6. As an advanced topic, the text then describes how to use Python and Tkinter to create graphical user interfaces (GUIs) in §7. Python's role in general scientific computing is described as a topic for further exploration (§8), as is the role of software licensing (§9) and project management via version control systems (§10). Exercises and examples occur throughout the text to concretely illustrate the language's usage and capabilities. A final project (§11) involves integrating several lessons from the text in order to address a structural bioinformatics question.

A collection of Supplemental Chapters (S1 Text) is also provided. The Chapters, which contain a few thousand lines of Python code, offer more detailed descriptions of much of the material in the main text. For instance, variables, functions and basic control flow are covered in Chapters 2, 3 and 5, respectively. Some topics are examined at greater depth, taking into account the interdependencies amongst topics—e.g., functions in Chaps 3, 7, and 13; lists, tuples, and other collections in Chaps 8, 9, and 10; OOP in Chaps 15 and 16. Finally, some topics that are either intermediate-level or otherwise not covered in the main text can be found in the Chapters, such as modules in Chap 4 and lambda expressions in Chap 13. The contents of the Chapters are summarized in Table 1 and in the S2 Text (§3, 'Sample Python Chapters').

**Table 1 Overview of the Supplemental Chapters**

| Chapter | Name | Topics |
|---|---|---|
| 00 | Setup | Commenting code. Running programs. Testing imports. |
| 01 | Introduction | Print function. Elementary arithmetic. Function definition syntax. Strings. |
| 02 | Variables | Variables and assignment. Semantics of `x=x+1` |
| 03 | Functions, I | Arguments and returns in functions. |
| 04 | Modules | Importing code from existing libraries. |
| 05 | Control Flow, I | `if` statements, indentation in Python. `%` and `//` operators. Boolean algebra. |
| 06 | Control Flow, II | Iteration using `for` and `while`. |
| 07 | Functions, II | Recursion and recursive problem decomposition. Collatz conjecture. |
| 08 | Collections, I | Syntax and semantics of sequences. Creation, modification, indexing, and slicing of tuples, lists, strings. Stack and heap storage. String formatting. |
| 09 | Collections, II | First-class functions. Techniques of list processing: `filter`, `fold`, `map`. |
| 10 | Collections, III | Syntax of dictionaries. Key-value pairs. |
| 11 | File I/O | Reading and writing files. Importance of leaving a file in a consistent state. |
| 12 | Graphics | Windows and basic graphics. Mouse events. External documentation. |
| 13 | Functions, III | Lambdas and functors. Nested function definitions and closures. Currying. |
| 14 | Exceptions | Indication of and recovery from errors. Exception hierarchy. |
| 15 | Classes, I | Object-oriented programming style; methods, members, inheritance. Assignment semantics. |
| 16 | Classes, II | Case studies of complex classes. Iterators. |
| 17 | Regexes | Essentials of regular expressions. |
| 18 | Tkinter, I | Creation of windows and widgets. Geometry managers. |
| 19 | Tkinter, II | Adding functionality to widgets. Tkinter variables. |

The Supplemental Chapters (S1 Text) consist of a few thousand lines of Python code, heavily annotated with explanatory material and covering the topics summarized here. In general, the material ranges from relatively basic to more intermediate and advanced levels as the Chapters progress. The latest versions of the Chapters are available at `http://p4b.muralab.org`.

## Using this Text

This text and the Supplemental Chapters work like the lecture and lab components of a course, and they are designed to be used in tandem. For readers who are new to programming, we suggest reading a section of text, including working through any examples or exercises in that section, and then completing the corresponding Supplemental Chapters before moving on to the next section; such readers should also begin by looking at §3.1 in the S2 Text, which describes how to interact with the Python interpreter, both in the context of a Unix Shell and in an integrated development environment (IDE) such as IDLE. For bioscientists who are somewhat familiar with a programming language (Python or otherwise), we suggest reading this text for background information and to understand the conventions used in the field, followed by a study of the Supplemental Chapters to learn the syntax of Python. For those with a strong programming background, this text will provide useful information about the software and conventions that commonly appear in the biosciences; the Supplemental Chapters will be rather familiar, in terms of algorithms and computer science fundamentals, while the biological examples and problems may be new for such readers.

### Typographic Conventions

The following typographic conventions appear in the remainder of this text: (i) all computer code is typeset in a `monospace` font; (ii) many terms are defined contextually, and are introduced in *italics*; (iii) **boldface** type is used for occasional emphasis; (iv) single (' ') and double (" ") quote marks are used either to indicate colloquial terms or else to demarcate character or word boundaries amidst the surrounding text (for clarity); (v) module names, filenames, pseudocode, and GUI-related strings appear as sans-serif text; and (vi) regular expressions are offset by `a gray background`, e.g. `.` denotes a period. We refer to delimiters in the text as (parentheses), [brackets], and {braces}.

Blocks of code are typeset in `monospace` font, with keywords in bold and strings in italics. Output appears on its own line without a line number, as in the following example:

```
1 if(True):
2   print("hello")
  hello
3 exit(0)
```

## Fundamentals of Programming

### Variables and Expressions

The concept of a variable offers a natural starting point for programming. A *variable* is a name that can be set to represent, or 'hold', a specific value. This definition closely parallels that found in mathematics. For example, the simple algebraic statement `x = 5` is interpreted mathematically as introducing the variable $x$ and assigning it the value 5. When Python encounters that same statement, the interpreter generates a variable named `x` (literally, by allocating memory), and assigns the value 5 to the variable name. The parallels between variables in Python and those in arithmetic continue in the following example, which can be typed at the prompt in any Python shell (§3.1 of the S2 Text describes how to access a Python shell):

```
1 x = 5
2 y = 7
3 z = x + 2 * y
4 print(z)
  19
```

As may be expected, the value of `z` is set equal to the sum of `x` and `2*y`, or in this case 19. The `print()` function makes Python output some text (the *argument*) to the screen; its name is a relic of early computing, when computers communicated with human users via ink-on-paper printouts. Beyond addition (+) and multiplication (*), Python can perform subtraction (−) and division (/) operations. Python is also natively capable (i.e., without add-on libraries) of other mathematical operations, including those summarized in Table 2.

To expand on the above example we will now use the math module, which is provided by default in Python. A *module* is a self-contained collection of Python code that can be imported,

**Table 2 Common Mathematical Operators in Python**

| Symbol | Functionality |
|---|---|
| + | addition |
| − | subtraction |
| * | multiplication |
| / | division |
| % | modulo (yields remainder after division) |
| // | integer division (truncates toward zero) |
| ** | exponentiation |
| `abs(a)` | absolute value of the number $a$, $|a|$ |
| `math.sin(x)` | sine of $x$ radians (other trigonometric functions are also available) |
| `math.factorial(n)` | factorial of $n$, $n!$ |
| `math.log(a,b)` | $log_b(a)$ (defaults to natural logarithm, if no base $b$ specified) |
| `math.sqrt(x)` | square root of $x$, $\sqrt{x}$ |

Common mathematical operators that are provided as built-in Python functions. Note that the behavior of / differs in versions of Python prior to 3.0; it previously acted as // does in recent versions of Python.

via the `import` command, into any other Python program in order to provide some functionality to the runtime environment. (For instance, modules exist to parse protein sequence files, read PDB files or simulation trajectories, compute geometric properties, and so on. Much of Python's *extensibility* stems from the ability to use (and write) various modules, as presented in Supplemental Chapter 4 (ch04modules.py).) A collection of useful modules known as the *standard library* is bundled with Python, and can be relied upon as always being available to a Python program. Python's math module (in the standard library) introduces several mathematical capabilities, including one that is used in this section: `sin()`, which takes an angle in radians and outputs the sine of that angle. For example,

```
1 import math
2 x = 21
3 y = math.sin(x)
4 print(y)
  0.8366556385360561
```

In the above program, the sine of `21 rad` is calculated, stored in `y`, and printed to the screen as the code's sole output. As in mathematics, an *expression* is formally defined as a unit of code that yields a value upon evaluation. As such, `x + 2*y`, `5 + 3`, `sin(pi)`, and even the number `5` alone, are examples of expressions (the final example is also known as a *literal*). All variable definitions involve setting a variable name equal to an expression.

Python's operator precedence rules mirror those in mathematics. For instance, `2+5*3` is interpreted as `2+(5*3)`. Python supports some operations that are not often found in arithmetic, such as `|` and `is`; a complete listing can be found in the official documentation [60]. Even complex expressions, like `x+3>>1|y&4>=5 or 6==z+ x)`, are fully (unambiguously) resolved by Python's operator precedence. However, few programmers would have the patience to determine the meaning of such an expression by simple inspection. Instead, when expressions become complex, it is almost always a good idea to use parentheses to explicitly clarify the order: `(((x+3 >> 1) | y&4) >= 5) or (6 == (z + x))`.

The following block reveals an interesting deviation from the behavior of a variable as typically encountered in mathematics:

```
1 x = 5
2 x = 2
3 print(x)
```

```
2
```

Viewed algebraically, the first two statements define an inconsistent system of equations (one with no solution) and may seem nonsensical. However, in Python, lines 1-2 are a perfectly valid pair of statements. When run, the `print` statement will display 2 on the screen. This occurs because Python, like most other languages, takes the statement `x = 2` to be a command to assign the value of 2 to `x`, ignoring any previous state of the variable `x`; such variable assignment statements are often denoted with the typographic convention '$x \leftarrow 2$'. Lines 1-2 above are **instructions** to the Python interpreter, rather than some system of equations with no solutions for the variable $x$. This example also touches upon the fact that a Python variable is purely a **reference** to an object (For now, take an *object* to simply be an addressable chunk of memory, meaning it can have a value and be referenced by a variable; objects are further described in §4.), such as the integer 5. This is a property of Python's *type system*. Python is said to be *dynamically typed*, versus *statically typed* languages such as C. In statically typed languages, a program's data (variable names) are bound to both an object and a type, and type checking is performed at compile-time; in contrast, variable names in a program written in a dynamically typed language are bound only to objects, and type checking is performed at run-time. An extensive treatment of this topic can be found in [61]. Dynamic typing is illustrated by the following example (The pound sign, #, starts a *comment*; Python ignores anything after a # sign, so in-line comments offer a useful mechanism for explaining and documenting one's code.):

```
1 x = 1           # x points to an integer object
2 y = x           # y points to the same object as does x (here, the integer 1)
3 x = "monty"     # x points to a different object now (a string)...
4 print(x)
monty
5 print(y)        # ... but that does not alter y's reference to its object
1
```

The above behavior results from the fact that, in Python, the notion of *type* (defined below) is attached to an object, not to any one of the potentially multiple names (variables) that reference that object. The first two lines illustrate that two or more variables can reference the same object (known as a *shared reference*), which in this case is of type `int`. When `y = x` is executed, `y` points to the object `x` points to (the integer 1). When `x` is changed, `y` still points to that original integer object. Note that Python strings and integers are *immutable*, meaning they cannot be changed in-place. However, some other object types, such as lists (described below), are mutable. These aspects of the language can become rather subtle, and the various features of the variable/object relationship—shared references, object mutability, etc.—can give rise to complicated scenarios. Supplemental Chapter 8 explores the Python memory model in more detail.

### Statements and Types

A *statement* is a command that instructs the Python interpreter to **do** something. All expressions are statements, but a statement need not be an expression. For instance, a statement that, upon execution, causes a program to stop running would never return a value, so it cannot be an expression. Most broadly, statements are instructions, while expressions are combinations of symbols (variables, literals, operators, etc.) that evaluate to a particular value. This particular value might be numerical (e.g., 5), a string (e.g., `'foo'`), Boolean (`True`/`False`), or some other type. Further distinctions between expressions and statements can become esoteric, and are not pertinent to much of the practical programming done in the biosciences.

The *type* of an object determines how the interpreter will treat the object when it is used. Given the code `x = 5`, we can say that "`x` is a variable that refers to an object that is of type int". We may simplify this to say "x is an int"; while technically incorrect, that is a shorter and more natural phrase. When the Python interpreter encounters the expression x + y, if x and y are [variables that point to objects of type] int, then the interpreter would use the addition hardware on the computer to add them. If, on the other hand, x and y were of type str, then Python would join them together. If one is a str and one is an int, the Python interpreter would "raise an exception" and the program would crash. Thus far, each variable we have encountered has been an integer (int) type, a string (str), or, in the case of sin()'s output, a real number stored to high precision (a float, for floating-point number). Strings and their constituent characters are among the most useful of Python's built-in types. Strings are sequences of characters, such as any word in the English language. In Python, a character is simply a string of length one. Each character in a string has a corresponding index, starting from 0 and ranging to index n-1 for a string of $n$ characters. Fig. 1 diagrams the composition and some of the functionality of a string, and the following code-block demonstrates how to define and manipulate strings and characters:

```
1 x = "red"
2 y = "green"
3 z = "blue"
4 print(x + y + z)
  redgreenblue
5 a = x[1]
6 b = y[2]
7 c = z[3]
8 print(a + " " + b + " " + c)
  e e e
```

Here, three variables are created by assignment to three corresponding strings. The first print may seem unusual: the Python interpreter is instructed to 'add' three strings; the interpreter joins them together in an operation known as *concatenation*. The second portion of code stores the character 'e', as extracted from each of the first three strings, in the respective variables, a, b and c. Then, their content is printed, just as the first three strings were. Note that spacing is not implicitly handled by Python (or most languages) so as to produce human-readable text; therefore, quoted whitespace was explicitly included between the strings (line 8; see also the underscore characters, '_', in Fig. 1).

   FIGURE 1 NEAR HERE

**Fig. 1 Strings in Python: Anatomy and Basic Behavior**. The anatomy and basic behavior of Python strings are shown, as samples of actual code (left panel) and corresponding conceptual diagrams (right panel). The Python interpreter prompts for user input on lines beginning with >>> (leftmost edge), while a starting ... denotes a continuation of the previous line; output lines are not prefixed by an initial character (e.g., the fourth line in this example). Strings are simply character array objects (of type str), and a sample string-specific method (replace) is shown on line 3. As with ordinary lists, strings can be 'sliced' using the syntax shown here: the first list element to be included in the slice is indexed by start, and the last included element is at stop-1, with an optional stride of size step (defaults to one). Concatenation, via the + operator, is the joining of whole strings or subsets of strings that are generated via slicing (as in this case). For clarity, the integer indices of the string positions are shown only in the forward (left to right) direction for mySnake1 and in the reverse direction for mySnake2. These two strings are sliced and concatenated to yield the object newSnake; note that slicing mySnake1 as [0:7] and not [0:6] means that a whitespace char is included between the two words in the resultant newSnake, thus obviating the need for further manipulations to insert whitespace (e.g., concatenations of the form word1+' '+word2).

**Exercise 1**: Write a program to convert a temperature in degrees Fahrenheit to degrees Celsius and Kelvin. The topic of user input has not been covered yet (to be addressed in §5), so begin with a variable that you pre-set to the initial temperature (in °F). Your code should convert the temperature to these other units and `print` it to the console.

## Functions

A deep benefit of the programming approach to problem-solving is that computers enable mechanization of repetitive tasks, such as those associated with data-analysis workflows. This is true in biological research and beyond. To achieve automation, a discrete and well-defined component of the problem-solving logic is encapsulated as a function. A *function* is a block of code that expresses the solution to a small, standalone problem/task; quite literally, a function can be any block of code that is defined by the user as being a function. Other parts of a program can then *call* the function to perform its task and possibly return a solution. For instance, a function can be repetitively applied to a series of input values via looping constructs (described below) as part of a data-processing pipeline.

Much of a program's versatility stems from its functions—the behavior and properties of each individual function, as well as the program's overall repertoire of available functions. Most simply, a function typically takes some values as its input *arguments* and acts on them; however, note that functions can be defined so as to not require any arguments (e.g., `print()` will give an empty line). Often, a function's arguments are specified simply by their position in the ordered list of arguments; e.g., the function is written such that the first expected argument is height, the second is weight, etc. As an alternative to such a system of *positional arguments*, Python has a useful feature called *keyword arguments*, whereby one can name a function's arguments and provide them in any order, e.g. `plotData(dataset=dats, color='red', width=10)`. Many scientific packages make extensive use of keyword arguments [62, 63]. The arguments can be variables, explicitly specified values (constants, string literals, etc.), or even other functions. Most generally, **any** expression can serve as an argument (Supplemental Chapter 13 covers more advanced usage, such as function objects). Evaluating a function results in its *return value*. In this way, a function's arguments can be considered to be its domain and its return values to be its range, as for any mathematical function $f$ that maps a domain $X$ to the range $Y$, $X \xrightarrow{f} Y$. If a Python function is given arguments outside its domain, it may return an invalid/nonsensical result, or even crash the program being run. The following illustrates how to define and then call (*invoke*) a function:

```
1 def myFun(a,b):
2   c = a + b
3   d = a - b
4   return c*d   # NB: a return does not 'print' anything on its own
5 x = myFun(1,3) + myFun(2,8) + myFun(-1,18)
6 print(x)
 -391
```

To see the utility of functions, consider how much code would be required to calculate `x` (line 5) in the absence of any calls to `myFun`. Note that discrete chunks of code, such as the body of a function, are delimited in Python via whitespace, not curly braces, {}, as in C or Perl. In Python, each level of indentation of the source code corresponds to a separate *block* of statements that group together in terms of program logic. The first line of above code illustrates the syntax to declare a function: A function definition begins with the keyword `def`, the following word names the function, and then the names within parentheses (separated by commas) define the arguments to the function. Finally, a colon terminates the function definition. (Default values of arguments can be specified as part of the function definition; e.g., writing line 1 as `def myFun(a=1,b=3):` would set default values of `a` and `b`.) The three statements after `def myFun(a,b):` are indented by some number of spaces (two, in this example), and so these three lines (2-4) constitute a *block*. In this block, lines 2-3 perform arithmetic operations on the arguments, and the final line of this function specifies the return value as the product of variables c and d. In effect, a `return` statement is what the function evaluates to when called, this return value taking the place of the original function call. It is also possible that a function returns nothing at all; e.g., a function might be intended to perform various manipulations and not necessarily return any output for downstream processing. For example, the following code defines (and then calls) a function that simply prints the values of three variables, without a `return` statement:

```
1 def readOut(a,b,c):
2   print("Variable 1 is: ", a)
3   print("Variable 2 is: ", b)
4   print("Variable 3 is: ", c)
5 readOut(1,2,4)
  Variable 1 is:  1
  Variable 2 is:  2
  Variable 3 is:  4
6 readOut(21,5553,3.33)
  Variable 1 is:  21
  Variable 2 is:  5553
  Variable 3 is:  3.33
```

## Code Organization and Scope

Beyond automation, structuring a program into functions also aids the modularity and interpretability of one's code, and ultimately facilitates the debugging process—an important consideration in all programming projects, large or small.

Python functions can be *nested*; that is, one function can be defined inside another. If a particular function is needed in only one place, it can be defined where it is needed and it will be unavailable elsewhere, where it would not be useful. Additionally, nested function definitions have access to the variables that are available when the nested function is defined. Supplemental Chapter 13 explores nested functions in greater detail. A function is an object in Python, just like a string or an integer. (Languages that allow function names to behave as objects are said to have 'first-class functions'.) Therefore, a function can itself serve as an argument to another function, analogous to the mathematical composition of two functions, $g(f(x))$. This property of the language enables many interesting programming techniques, as explored in Supplemental Chapters 9 and 13.

A variable created inside a block, e.g. within a function, cannot be accessed by name from outside that block. The variable's *scope* is limited to the block wherein it was defined. A variable or function that is defined outside of every other block is said to be *global* in scope. Variables can appear within the scope in which they are defined, or any block within that scope, but the reverse is not true: variables cannot escape their scope. This rule hierarchy is diagrammed in Fig. 2. There is only one global scope, and variables in it necessarily 'persist' between function calls (unlike variables in local scope). For instance, consider two functions, fun1 and fun2; for convenience, denote their local scopes as $\ell_1$ and $\ell_2$, and denote the global scope as $\mathcal{G}$. Starting in $\mathcal{G}$, a call to fun1 places us in scope $\ell_1$. When fun1 successfully returns, we return to scope $\mathcal{G}$; a call to fun2 places us in scope $\ell_2$, and after it completes we return yet again to $\mathcal{G}$. We always return to $\mathcal{G}$. In this sense, local scope varies whereas global scope (by definition) persists between function calls, is available inside/outside of functions, etc. Explicitly tracking the precise scope of every object in a large body of code can be cumbersome. However, this is rarely burdensome in practice: Variables are generally defined (and are therefore in scope) where they are used. After encountering some out-of-scope errors, and gaining experience with nested functions and variables, carefully managing scope in a consistent and efficient manner will become an implicit skill (and will be reflected in one's coding 'style').

    FIGURE 2 NEAR HERE

**Fig. 2 Python's Scope Hierarchy and Variable Name Resolution**. As described in the text, multiple names (variables) can reference a single object. Conversely, can a single variable, say `x`, reference multiple objects in a unique and well-defined manner? Exactly this is enabled by the concept of a *namespace*, which can be viewed as the set of all name↔object mappings for all variable names and objects at a particular 'level' in a program. This is a crucial concept, as everything in Python is an object. The key idea is that name↔object mappings are insulated from one another, and therefore free to vary, at different 'levels' in a program—e.g., `x` might refer to object `obj2` in a block of code buried (many indentation levels deep) within a program, whereas the same variable name `x` may reference an entirely different object, `obj1`, when it appears as a top-level (module-level) name definition. This seeming ambiguity is resolved by the notion of variable scope. The term *scope* refers to the level in the namespace hierarchy that is searched for name↔object mappings; different mappings can exist in different scopes, thus avoiding potential name collisions. At a specific point in a block of code, in what order does Python search the namespace levels? (And, which of the potentially multiple name↔object mappings takes precedence?) Python resolves variable names by traversing scope in the order L→E→G→B, as shown here. L stands for the **l**ocal, innermost scope, which contains local names and is searched first; E follows, and is the scope of any **e**nclosing functions; next is G, which is the namespace of all **g**lobal names in the currently loaded modules; finally, the outermost scope B, which consists of Python's **b**uilt-in names (e.g., `int`), is searched last. The two code examples in this figure demonstrate variable name resolution at local and global scope levels. In the code on the right-hand side, the variable `e` is used both (i) as a name imported from the math module (global scope) and (ii) as a name that is local to a function body, albeit with the `global` keyword prior to being assigned to the integer `-1234`. This construct leads to a confusing flow of logic (colored arrows), and is considered poor programming practice.

Well-established practices have evolved for structuring code in a logically organized (often hierarchical) and 'clean' (lucid) manner, and comprehensive treatments of both practical and abstract topics are available in numerous texts. See, for instance, the practical guide *Code Complete* [64], the intermediate-level *Design Patterns: Elements of Reusable Object-Oriented Software* [65], and the classic (and more abstract) texts *Structure and Interpretation of Computer Programs* [39] and *Algorithms* [50]; a recent, and free, text in the latter class is *Introduction to Computing* [40]. Another important aspect of coding is closely related to the above: Usage of brief, yet informative, names as identifiers for variables and function definitions. Even a mid-sized programming project can quickly grow to thousands of lines of code, employ hundreds of functions, and involve hundreds of variables. Though the fact that many variables will lie outside the scope of one another lessens the likelihood of undesirable references to ambiguous variable names, one should note that careless, inconsistent, or undisciplined nomenclature **will** confuse later efforts to understand a piece of code, for instance by a collaborator or, after some time, even the original programmer. Writing clear, well-defined and well-annotated code is an essential skill to develop. Table 3 outlines some suggested naming practices.

Python minimizes the problems of conflicting names via the concept of namespaces. A *namespace* is the set of all possible (valid) names that can be used to uniquely identify an object at a given level of scope, and in this way it is a more generalized concept than scope (see also Fig. 2). To access a name in a different namespace, the programmer must tell the interpreter what namespace to search for the name. An `import`ed module, for example,

**Table 3 Sample Variable-naming Schemes in Python**

| Variable Name | Valid? | Comments |
|---|---|---|
| `numHelix` | yes | By using the camelCase convention (capitalize the first letter of every word following the first), this variable name is reasonably descriptive, easily read, and brief. |
| `123myVar!?` | no | This variable uses forbidden characters ('!' and '?'), and it also starts with digits; all special characters apart from '_' are disallowed, and no variable can begin with a digit. |
| `int` | yes | This Python built-in function is technically allowed as a variable name; however, its attempted usage may yield unexpected behavior, as the statement '`int(x)`' converts an object '`x`' to type `int`. |
| `else` | no | This is an element of the basic syntax in Python (a *keyword*). Using a keyword as a variable name is a syntax error. Other common keywords are `for`, `if`, `not`, `return`, and `def`. |
| `_myVar2` | yes | Though fairly nondescript, this variable does not contain forbidden characters and is technically valid (note that the underscore character is generally used to denote [protected] member variables in objects). |
| `IlI1lI` | yes | While technically valid, such a name is unnecessarily frustrating for those reading and maintaining code. |

These examples of variable-naming schemes in Python are annotated with comments as to their validity and their suitability (the latter is more subjective). In practice, variables at higher (broader) levels of scope tend to have names that are longer, more descriptive, and less ambiguous than at lower (narrower) levels of scope.

creates its own new namespace. The `math` module creates a namespace (called `math`) that contains the `sin()` function. To access `sin()`, the programmer must qualify the function call with the namespace to search, as in `y = math.sin(x)`. This precision is necessary because merging two namespaces that might possibly contain the same names (in this case, the `math` namespace and the global namespace) results in a *name collision*. Another example would be to consider the files in a Unix directory (or a Windows folder); in the namespace of this top-level directory, one file can be named foo1 and another foo2, but there cannot be two files named foo—that would be a name collision.

**Exercise 2**: Recall the temperature conversion program of Exercise 1. Now, write a function to perform the temperature conversion; this function should take one argument (the input temperature). To test your code, use the function to convert and print the output for some arbitrary temperatures of your choosing.

## Control Flow: Conditionals

> "Begin at the beginning," the King said gravely, "and go on till you come to the end; then, stop."
>
> — Lewis Carroll, *Alice in Wonderland*

Thus far, all of our sample code and exercises have featured a linear flow, with statements executed and values emitted in a predictable, deterministic manner. However, most scientific datasets are not amenable to analysis via a simple, predefined stream of instructions. For example, the initial data-processing stages in many types of experimental pipelines may entail the assignment of statistical confidence/reliability scores to the data, and then some form of decision-making logic might be applied to filter the data. Often, **if** a particular datum does not meet some statistical criterion and is considered a likely outlier, **then** a special task is performed; **otherwise**, another (default) route is taken. This branched **if**–**then**–**else** logic is a key decision-making component of virtually any algorithm, and it exemplifies the concept of control flow. The term *control flow* refers to the progression of logic as the Python interpreter traverses the code and the program 'runs'—transitioning, as it runs, from one state to the next, choosing which statements are executed, iterating over a loop some number of times, and so on. (Loosely, the *state* can be taken as the line of code that is being executed, along with the collection of all variables, and their values, accessible to a running program at any instant;

given the precise state, the next state of a deterministic program can be predicted with perfect precision.) The following code introduces the `if` statement:

```
1 from random import randint
2 a = randint(0,100)   # get a random integer between 0 and 100 (inclusive )
3 if(a < 50):
4   print("variable is less than 50")
5 else:
6   print("the variable is not less than 50")
  variable is less than 50
```

In this example, a random integer between 0 and 100 is assigned to the variable `a`. (Though not applicable to `randint`, note that many sequence/list-related functions, such as `range(a,b)`, generate collections that start at the first argument and end just before the last argument. This is because the function `whatEver(a,b)` produces $b - a$ items starting at $a$; with a default stepsize of one, this makes the endpoint `b-1`.) Next, the `if` statement tests whether the variable is less than 50. If that condition is unfulfilled, the block following `else` is executed. Syntactically, `if` is immediately followed by a *test condition*, and then a colon to denote the start of the `if` statement's block (Fig. 3 illustrates the use of conditionals). Just as with functions, the further indentation on line 4 creates a block of statements that are executed together (here, the block has only one statement). Note that an `if` statement can be defined without a corresponding `else` block; in that case, Python simply continues executing the code that is indented by one less level (i.e., at the same indentation level as the `if` line). Also, Python offers a built-in `elif` keyword (a contraction of 'else if') that tests a subsequent conditional if and only if the first condition is not met. A series of `elif` statements can be used to achieve similar effects as the `switch`/`case` statement constructs found in C and in other languages (including Unix shell scripts) that are often encountered in bioinformatics.

   FIGURE 3 NEAR HERE

**Fig. 3 Sample Flowchart for a Sorting Algorithm**. This flowchart illustrates the conditional constructs, loops, and other elements of control flow that comprise an algorithm for sorting, from smallest to largest, an arbitrary list of numbers (the algorithm is known as 'bubble sort'). In this type of diagram, arrows symbolize the flow of logic (control flow), rounded rectangles mark the start and end points, slanted parallelograms indicate I/O (e.g., a user-provided list), rectangles indicate specific subroutines or procedures (blocks of statements), and diamonds denote conditional constructs (branch points). Note that this sorting algorithm involves a pair of nested loops over the list size (blue and orange), meaning that the calculation cost will go as the square of the input size (here, an $N$-element list); this cost can be halved by adjusting the inner loop conditional to be "j < N – i – 1", as the largest $i$ elements will have already reached their final positions.

   Now, consider the following extension to the above block of code. Is there any fundamental issue with it?

```
1 from random import randint
2 a = randint(0,100)
3 if(a < 50):
4   print("variable is less than 50")
5 if(a > 50):
6   print("variable is greater than 50")
7 else:
8   print("the variable must be 50")
  variable is greater than 50
```

This code will function as expected for `a=50`, as well as values exceeding `50`. However, for `a` less than `50`, the print statements will be executed from **both** the less-than (line 4) and equal-to (line 8) comparisons. This erroneous behavior results because an `else` statement is bound solely to the `if` statement that it directly follows; in the above code-block, an `elif` would have been the appropriate keyword for line 5. This example also underscores the danger of assuming that lack of a certain condition (a `False` built-in Boolean type) necessarily implies the fulfillment of a second condition (a `True`) for comparisons that seem, at least superficially, to be linked. In writing code with complicated streams of logic (conditionals and beyond), robust and somewhat redundant logical tests can be used to mitigate errors and unwanted behavior. A strategy for building streams of conditional statements into code, and for debugging existing codebases, involves (i) outlining the range of possible inputs (and their expected outputs), (ii) crafting the code itself, and then (iii) testing each possible type of input, carefully tracing the logical flow executed by the algorithm against what was originally anticipated. In step (iii), a careful examination of 'edge cases' can help debug code and pinpoint errors or unexpected behavior. (In software engineering parlance, *edge cases* refer to extreme values of parameters, such as minima/maxima when considering ranges of numerical types. Recognition of edge-case behavior is useful, as a disproportionate share of errors occur near these cases; for instance, division by zero can crash a function if the denominator in each division operation that appears in the function is not carefully checked and handled appropriately. Though beyond the scope of this primer, note that Python supplies powerful error-reporting and exception-handling capabilities; see, for instance, *Python Programming* [66] for more information.) Supplemental Chapters 14 and 16 provide detailed examples of testing the behavior of code.

**Exercise 3**: Recall the temperature-conversion program designed in Exercises 1 and 2. Now, rewrite this code such that it accepts two arguments: The initial temperature, and a letter designating the units of that temperature. Have the function convert the input temperature to the alternative scale. If the second argument is `'C'`, convert the temperature to Fahrenheit, if that argument is `'F'`, convert it to Celsius.

Integrating what has been described thus far, the following example demonstrates the power of control flow—not just to define computations in a structured/ordered manner, but also to solve real problems by devising an algorithm. In this example, we sort three randomly chosen integers:

```
1  from random import randint
2  def numberSort():
3    a = randint(0,100)
4    b = randint(0,100)
5    c = randint(0,100)
6    # reminder: text following the pound sign on one line is a comment in Python.
7    # begin sort; note the nested conditionals here
8    if ((a > b) and (a > c)):
9      largest = a
10     if(b > c):
11       second = b
12       third = c
13     else:
14       second = c
15       third = b
16   # a must not be largest
17   elif(b > c):
18     largest = b
19     if(c > a):
20       second = c
```

```
21         third = a
22      else:
23         second = a
24         third = c
25   # a and b are not largest , thus c must be
26   else:
27      largest = c
28      if(b < a):
29         second = a
30         third = b
31      else:
32         second = b
33         third = a
34   # Python's assert  function  can be used for sanity  checks.
35   # If the argument to assert () is False, the program will crash.
36   assert(largest > second)
37   assert(second > third)
38   print("Sorted: ", largest, ", ", second, ", ", third)
39 numberSort()
  Sorted:  50 ,  47 ,  11
```

## Control Flow: Repetition via While Loops

Whereas the `if` statement tests a condition exactly once and branches the code execution accordingly, the `while` statement instructs an enclosed block of code to repeat so long as the given condition (the *continuation condition*) is satisfied. In fact, `while` can be considered as a repeated `if`. This is the simplest form of a loop, and is termed a *while loop* (Fig. 3). The condition check occurs once before entering the associated block; thus, Python's `while` is a *pre-test* loop. (Some languages feature looping constructs wherein the condition check is performed after a first iteration; C's *do–while* is an example of such a *post-test* loop. This is mentioned because looping constructs should be carefully examined when comparing source code in different languages.) If the condition is true, the block is executed and then the interpreter effectively jumps to the `while` statement that began the block. If the condition is false, the block is skipped and the interpreter jumps to the first statement after the block. The code below is a simple example of a while loop, used to generate a counter that prints each integer between `1` and `100` (inclusive):

```
1 counter = 1
2 while(counter <= 100):
3    print(counter)
4    counter = counter + 1
  1
  2
  ...
  99
  100
5 print("done!")
  done!
```

This code will begin with a variable, then print and increment it until its value is `101`, at which point the enclosing while loop ends and a final string (line 5) is printed. Crucially, one should verify that the loop termination condition can, in fact, be reached. If not—e.g., if the loop were specified as `while(True):` for some reason—then the loop would continue indefinitely, creating an infinite loop that would render the program unresponsive. (In many environments, such as a Unix shell, the keystroke `ctrl-c` can be used as a *keyboard interrupt* to break out of the loop.)

**Exercise 4**: With the above example as a starting point, write a function that chooses two randomly-generated integers between 0 and 100, inclusive, and then prints all numbers between these two values, counting from the lower number to the upper number.

## Recursion

> "In order to understand recursion, you must first understand recursion."
> 
> — Anonymous

Recursion is a subtle concept. A while loop is conceptually straightforward: A block of statements comprising the body of the loop is repeatedly executed as long as a condition is true. A *recursive* function, on the other hand, calls **itself** repeatedly, effectively creating a loop. Recursion is the most natural programming approach (or *paradigm*) for solving a complex problem that can be decomposed into (easier) subproblems, each of which resembles the overall problem. Mathematically, problems that are formulated in this manner are known as *recurrence relations*, and a classic example is the factorial (below). Recursion is so fundamental and general a concept that iterative constructs (for, while loops) can be expressed recursively; in fact, some languages dispense with loops entirely and rely on recursion for all repetition. The key idea is that a recursive function calls itself from within its own function body, thus progressing one step closer to the final solution at each self-call. The recursion terminates once it has reached a trivially simple final operation, termed the *base case*. (Here, the word 'simple' means only that evaluation of the final operation yields no further recursive steps, with no implication as to the computational complexity of that final operation.) Calculation of the factorial function, $f(n) = n!$, is a classic example of a problem that is elegantly coded in a recursive manner. Recall that the factorial of a natural number, $n$, is defined as:

$$n! = \begin{cases} 1 & n = 1 \quad \text{(base case)} \\ n * (n-1)! & n > 1 \end{cases} \quad (1)$$

This function can be compactly implemented in Python like so:

```python
def factorial(n):
    assert(n > 0)  #Crash on invalid input
    if(n == 1):
        return 1
    else:
        return n * factorial(n-1)
```

A call to this `factorial` function will return 1 if the input is equal to one, and otherwise will return the input value multiplied by the factorial of that integer less one (`factorial(n-1)`). Note that this recursive implementation of the factorial perfectly matches its mathematical definition. This often holds true, and many mathematical operations on data are most easily expressed recursively. When the Python interpreter encounters the call to the `factorial` function within the function block itself (line 6), it generates a new instance of the function on the fly, while retaining the original function in memory (technically, these function instances occupy the runtime's *call stack*). Python places the current function call on hold in the call stack while the newly-called function is evaluated. This process continues until the base case is reached, at which point the function returns a value. Next, the previous function instance in the call stack resumes execution, calculates its result, and returns it. This process of traversing the call stack continues until the very first invocation has returned. At that point, the call stack is empty and the function evaluation has completed.

## Expressing Problems Recursively

Defining recursion simply as a function calling itself misses some nuances of the recursive approach to problem-solving. Any difficult problem (e.g., $f(n) = n!$) that can be expressed as a simpler instance of the same problem (e.g., $f(n) = n * f(n-1)$) is amenable to a recursive solution. Only when the problem is trivially easy (1!, `factorial(1)` above) does the recursive solution give a direct (one-step) answer. Recursive approaches fundamentally differ from more iterative (also known as *procedural*) strategies: Iterative constructs (loops) express the entire solution to a problem in more explicit form, whereas recursion repeatedly makes a problem simpler until it is trivial. Many data-processing functions are most naturally and compactly solved via recursion.

The recursive descent/ascent behavior described above is extremely powerful, and care is required to avoid pitfalls and frustration. For example, consider the following addition algorithm, which uses the equality operator (==) to test for the base case:

```
def badRecursiveAdder(x):
  if(x == 1):
    return x
  else:
    return x + badRecursiveAdder(x-2)
```

This function does include a base case (lines 2-3), and at first glance may seem to act as expected, yielding a sequence of squares $(1, 4, 9, 16. \ldots)$ for $x = 1, 3, 5, 7, \ldots$. Indeed, for odd `x` greater than `1`, the function will behave as anticipated. However, if the argument is negative or is an even number, the base case will never be reached (note that line 5 subtracts `2`), causing the function call to simply hang, as would an infinite loop. (In this scenario, Python's *maximum recursion depth* will be reached and the call stack will overflow.) Thus, in addition to defining the function's base case, it is also crucial to confirm that **all** possible inputs will reach the base case. A valid recursive function must progress towards—and eventually reach—the base case with every call. More information on recursion can be found in Supplemental Chapter 7, in Chapter 4 of [40], and in most computer science texts.

**Exercise 5**: Consider the Fibonacci sequence of integers, $0, 1, 1, 2, 3, 5, 8, 13, \ldots$, given by

$$F_n = \begin{cases} n & n \leq 1 \\ F_{n-1} + F_{n-2} & n > 1 \end{cases} \quad (2)$$

This sequence appears in the study of phyllotaxis and other areas of biological pattern formation (see, e.g., [67]). Now, write a recursive Python function to compute the $n^{\text{th}}$ Fibonacci number, $F_n$, and test that your program works. Include an `assert` to make sure the argument is positive. Can you generalize your code to allow for different seed values ($F_0 = l$, $F_1 = m$, for integers $l$ and $m$) as arguments to your function, thereby creating new sequences? (Doing so gets you one step closer to Lucas sequences, $L_n$, which are a highly general class of recurrence relations.)

**Exercise 6**: Many functions can be coded both recursively and iteratively (using loops), though often it will be clear that one approach is better suited to the given problem (the factorial is one such example). In this exercise, devise an **iterative** Python function to compute the factorial of a user-specified integer argument. As a bonus exercise, try coding the Fibonacci sequence in iterative form. Is this as straightforward as the recursive approach? Note that Supplemental Chapter 7 might be useful here.

## Data Collections: Tuples, Lists, for Loops, and Dictionaries

A staggering degree of algorithmic complexity is possible using only variables, functions, and control flow concepts. However, thus far, numbers and strings are the only data types that have been discussed. Such data types can be used to represent protein sequences (a string) and molecular masses (a floating point number), but actual scientific data are seldom so simple!

The data from a mass spectrometry experiment are a list of intensities at various $m/z$ values (the mass spectrum). Optical microscopy experiments yield thousands of images, each consisting of a large two-dimensional array of pixels, and each pixel has color information that one may wish to access [68]. A protein multiple sequence alignment can be considered as a two-dimensional array of characters drawn from a 21-letter alphabet (one letter per amino acid (AA) and a gap symbol), and a protein 3D structural alignment is even more complex. Phylogenetic trees consist of sets of species, individual proteins, or other taxonomic entities, organized as (typically) binary trees with branch weights that represent some metric of evolutionary distance. A trajectory from an MD or Brownian dynamics simulation is especially dense: Cartesian coordinates and velocities are specified for upwards of $10^6$ atoms at $>10^6$ time-points (every ps in a μs-scale trajectory). As illustrated by these examples, real scientific data exhibit a level of complexity far beyond Python's relatively simple built-in data types. Modern datasets are often quite heterogeneous, particularly in the biosciences [69], and therefore *data abstraction* and *integration* are often the major goals. The data challenges hold true at all levels, from individual RNA transcripts [70] to whole bacterial cells [71] to biomedical informatics [72].

In each of the above examples, the relevant data comprise a collection of entities, each of which, in turn, is of some simpler data type. This unifying principle offers a way forward. The term *data structure* refers to an object that stores data in a specifically organized (structured) manner, as defined by the programmer. Given an adequately well-specified/defined data structure, arbitrarily complex collections of data can be readily handled by Python, from a simple array of integers to a highly intricate, multi-dimensional, heterogeneous (mixed-type) data structure. Python offers several built-in sequence data structures, including strings, lists, and tuples.

## Tuples

A *tuple* (pronounced like 'couple') is simply an ordered sequence of objects, with essentially no restrictions as to the types of the objects. Thus, the tuple is especially useful in building data structures as higher-order collections. Data that are inherently sequential (e.g., time-series data recorded by an instrument) are naturally expressed as a tuple, as illustrated by the following syntactic form: `myTuple = (0,1,3)`. The tuple is surrounded by parentheses, and commas separate the individual elements. The empty tuple is denoted `()`, and a tuple of one element contains a comma after that element, e.g., `(1,)`; the final comma lets Python distinguish between a tuple and a mathematical operation. That is, `2*(3+1)` must not treat `(3+1)` as a tuple. A parenthesized expression is therefore not made into a tuple unless it contains commas. (The `type` function is a useful built-in function to probe an object's type. At the Python interpreter, try the statements `type((1))` and `type((1,))`. How do the results differ?)

A tuple can contain any sort of object, including another tuple. For example, `diverseTuple = (15.38,"someString",(0,1))` contains a floating-point number, a string, and another tuple. This versatility makes tuples an effective means of representing complex or heterogeneous data structures. Note that any component of a tuple can be referenced using the same notation used to index individual characters within a string; e.g., `diverseTuple[0]` gives `15.38`.

In general, data are optimally stored, analyzed, modified, and otherwise processed using data structures that reflect any underlying structure of the data itself. Thus, for example, two-dimensional datasets are most naturally stored as tuples of tuples. This abstraction can be taken to arbitrary depth, making tuples useful for storing arbitrarily complex data. For instance, tuples have been used to create generic tensor-like objects. These rich data structures have been used in developing new tools for the analysis of MD trajectories [18] and to represent biological sequence information as hierarchical, multidimensional entities that are amenable to further processing in Python [20].

As a concrete example, consider the problem of representing signal intensity data collected over time. If the data are sampled with perfect periodicity, say every second, then the information could be stored (most compactly) in a one-dimensional tuple, as a simple succession of intensities; the index of an element in the tuple maps to a time-point (index `0` corresponds to the measurement at time $t_0$, index `1` is at time $t_1$, etc.). What if the data were sampled unevenly in time? Then each datum could be represented as an ordered pair, $(t, I(t))$, of the intensity $I$ at each time-point $t$; the full time-series of measurements is then given by the sequence of 2-element tuples, like so:

```
dataSet = ((0.00,0.2),
           (0.17,0.3),
           (0.33,0.4),
           (0.40,0.2),
           (0.90,0.0))

print(dataSet[4])     # prints the final ordered pair
(0.9, 0.0)
print(dataSet[0][1])  # prints the intensity at timestep zero
0.2
print(dataSet[2][0])  # prints the time at timestep two
0.33
```

Three notes concern the above code: (i) From this two-dimensional data structure, the syntax `dataSet[i][j]` retrieves the $j^{th}$ element from the $i^{th}$ tuple. (ii) Negative indices can be used as shorthand to index from the end of most collections (tuples, lists, etc.), as shown in Fig. 1; thus, in the above example `dataSet[-1]` represents the same value as `dataSet[4]`. (iii) Recall that Python treats all lines of code that belong to the same block (or degree of indentation) as a single unit. In the example above, the first line alone is not a valid (closed) expression, and Python allows the expression to continue on to the next line; the lengthy `dataSet` expression was formatted as above in order to aid readability.

Once defined, a tuple cannot be altered; tuples are said to be *immutable* data structures. This rigidity can be helpful or restrictive, depending on the context and intended purpose. For instance, tuples are suitable for storing numerical constants, or for ordered collections that are generated once during execution and intended only for referencing thereafter (e.g., an input stream of raw data).

## Lists

A mutable data structure is the Python *list*. This built-in sequence type allows for the addition, removal, and modification of elements. The syntactic form used to define lists resembles the definition of a tuple, except that the parentheses are replaced with square brackets, e.g. `myList = [0,1,42,78]`. (A trailing comma is unnecessary in one-element lists, as `[1]` is unambiguously a list.) As suggested by the preceding line, the elements in a Python list are typically more homogeneous than might be found in a tuple: The statement `myList2=['a',1]`, which defines a list containing both string and numeric types, is technically valid, but `myList2=['a','b']` or `myList2=[0,1]` would be more frequently encountered in practice. Note that `myList[1]=3.14` is a perfectly valid statement that can be applied to the already-defined object named `myList` (as long as `myList` already contains two or more elements), resulting in the modification of the second element in the list. Finally, note that `myList[5]=3.14` will raise an error, as the list defined above does not contain a sixth element. The index is said to be *out of range*, and a valid approach would be to append the value via `myList.append(3.14)`.

The foregoing description only scratches the surface of Python's built-in data structures. Several functions and methods are available for lists, tuples, strings, and other built-in types.

For lists, `append`, `insert`, and `remove` are examples of oft-used methods; the function `len()` returns the number of items in a sequence or collection, such as the length of a string or number of elements in a list. All of these 'list methods' behave similarly as any other function—arguments are generally provided as input, some processing occurs, and values may be returned. (Section §4 elaborates the relationship between functions and methods.)

### Iteration with For Loops

Lists and tuples are examples of *iterable* types in Python, and the *for loop* is a useful construct in handling such objects. (Custom iterable types are introduced in Supplemental Chapter 17.) A Python `for` loop iterates over a collection, which is a common operation in virtually all data-analysis workflows. Recall that a `while` loop requires a counter to track progress through the iteration, and this counter is tested against the continuation condition. In contrast, a `for` loop handles the count implicitly, given an argument that is an iterable object:

```
myData = [1.414, 2.718, 3.142, 4.669]
total = 0
for datum in myData:
    # the next statement uses a compound assignment operator; in
    # the addition assignment operator, a += b means a = a + b
    total += datum
    print("added " + str(datum) + " to sum.")
    #str makes a string from datum so we can concatenate with +.
added 1.414 to sum.
added 2.718 to sum.
added 3.142 to sum.
added 4.669 to sum.
print(total)
11.942999999999998
```

In the above loop, all elements in `myData` are of the same type (namely, floating-point numbers). This is not mandatory. For instance, the heterogeneous object `myData=['a','b',1,2]` is iterable, and therefore it is a valid argument to a `for` loop (though not the above loop, as string and integer types cannot be mixed as operands to the + operator). The context dependence of the + symbol, meaning either numeric addition or a concatenation operator, depending on the arguments, is an example of *operator overloading*. (Together with dynamic typing, operator overloading helps make Python a highly expressive programming language.) In each iteration of the above loop, the variable `datum` is assigned each successive element in `myData`; specifying this iterative task as a `while` loop is possible, but less straightforward. Finally, note the syntactic difference between Python's `for` loops and the `for(<initialize>; <condition>; <update>) {<body>}` construct that is found in C, Perl, and other languages encountered in computational biology.

**Exercise 7**: Consider the fermentation of glucose into ethanol: $C_6H_{12}O_6 \rightarrow 2C_2H_5OH + 2CO_2$. A fermentor is initially charged with 10,000 liters of feed solution and the rate of carbon dioxide production is measured by a sensor in moles/hour. At $t=$10, 20, 30, 40, 50, 60, 70, and 80 hours, the $CO_2$ generation rates are 58.2, 65.2, 67.8, 65.4, 58.8, 49.6, 39.1, and 15.8 moles/hour respectively. Assuming that each reading represents the average $CO_2$ production rate over the previous ten hours, calculate the total amount of $CO_2$ generated and the final ethanol concentration in grams per liter. Note that Supplemental Chapters 6 and 9 might be useful here.

**Exercise 8**: Write a program to compute the distance, $d(\mathbf{r_1}, \mathbf{r_2})$, between two arbitrary (user-specified) points, $\mathbf{r_1} = (x_1, y_1, z_1)$ and $\mathbf{r_2} = (x_2, y_2, z_2)$, in 3D space. Use the usual Euclidean distance between two points—the straight-line, "as the bird flies" distance. Other distance metrics, such as the Mahalanobis and Manhattan distances, often appear in computational biology too. With your code in hand, note the ease with which you can adjust your entire data-analysis workflow simply by modifying a few lines of code that correspond to the definition of the distance function. As a bonus exercise, generalize your code to read in a list of points and compute the total path length. Supplemental Chapters 6, 7, and 9 might be useful here.

### Sets and Dictionaries

Whereas lists, tuples and strings are ordered (sequential) data types, Python's *sets* and *dictionaries* are unordered data containers. Dictionaries, also known as *associative arrays* or *hashes* in Perl and other common languages, consist of key:value pairs enclosed in braces. They are particularly useful data structures because, unlike lists and tuples, the values are not restricted to being indexed solely by the integers corresponding to sequential position in the data series. Rather, the keys in a dictionary serve as the index, and they can be of any immutable data type (strings, numbers, or tuples of immutable data). A simple example, indexing on three-letter abbreviations for amino acids and including molar masses, would be `aminoAcids={'ala':('a','alanine', 89.1),'cys':('c','cysteine', 121.2)}`. A dictionary's items are accessed via square brackets, analogously as for a tuple or list, e.g., `aminoAcids['ala']` would retrieve the tuple `('a','alanine', 89.1)`. As another example, dictionaries can be used to create lookup tables for the properties of a collection of closely related proteins. Each key could be set to a unique identifier for each protein, such as its UniProt ID (e.g., Q8ZYG5), and the corresponding values could be an intricate tuple data structure that contains the protein's isoelectric point, molecular weight, PDB accession code (if a structure exists), and so on. Dictionaries are described in greater detail in Supplemental Chapter 10.

### Further Data Structures: Trees and Beyond

Python's built-in data structures are made for sequential data, and using them for other purposes can quickly become awkward. Consider the task of representing genealogy: an individual may have some number of children, and each child may have their own children, and so on. There is no straightforward way to represent this type of information as a list or tuple. A better approach would be to represent each organism as a tuple containing its children. Each of those elements would, in turn, be another tuple with children, and so on. A specific organism would be a *node* in this data structure, with a *branch* leading to each of its *child* nodes; an organism having no children is effectively a *leaf*. A node that is not the child of any other node would be the *root* of this tree. This intuitive description corresponds, in fact, to exactly the terminology used by computer scientists in describing *trees* [73]. Trees are pervasive in computer science. This document, for example, could be represented purely as a list of characters, but doing so neglects its underlying structure, which is that of a tree (sections, sub-sections, sub-sub-sections, ...). The whole document is the root entity, each section is a node on a branch, each sub-section a branch from a section, and so on down through the paragraphs, sentences, words, and letters. A common and intuitive use of trees in bioinformatics is to represent phylogenetic relationships. However, trees are such a general data structure that they also find use, for instance, in computational geometry applications to biomolecules (e.g., to optimally partition data along different spatial dimensions [74, 75]).

Trees are, by definition, *(i) acyclic*, meaning that following a branch from node $i$ will never lead back to node $i$, and any node has exactly one parent; and *(ii) directed*, meaning that a node knows only about the nodes 'below' it, not the ones 'above' it. Relaxing these requirements gives a *graph* [76], which is an even more fundamental and universal data structure: A graph is a set of vertices that are connected by edges. Graphs can be subtle to work with and a number of clever algorithms are available to analyze them [77].

There are countless data structures available, and more are constantly being devised. Advanced examples range from the biologically-inspired neural network, which is essentially a graph wherein the vertices are linked into communication networks to emulate the neuronal layers in a brain [78], to very compact probabilistic data structures such as the Bloom filter [79], to self-balancing trees [80] that provide extremely fast insertion and removal of elements for performance-critical code, to copy-on-write B-trees that organize terabytes of information on hard drives [81].

# Object-oriented Programming in a Nutshell: Classes, Objects, Methods, & All That

## OOP in Theory: Some Basic Principles

Computer programs are characterized by two essential features [82]: (i) **algorithms** or, loosely, the 'programming logic', and (ii) **data structures**, or how data are represented within the program, whether certain components are manipulable, iterable, etc. The object-oriented programming (OOP) paradigm, to which Python is particularly well-suited, treats these two features of a program as inseparable. Several thorough treatments of OOP are available, including texts that are independent of any language [83] and books that specifically focus on OOP in Python [84]. The core ideas are explored in this section and in Supplemental Chapters 15 and 16.

Most scientific data have some form of inherent structure, and this serves as a starting point in understanding OOP. For instance, the time-series example mentioned above is structured as a series of ordered pairs, $(t, I(t))$, an X-ray diffraction pattern consists of a collection of intensities that are indexed by integer triples $(h, k, l)$, and so on. In general, the intrinsic structure of scientific data cannot be easily or efficiently described using one of Python's standard data structures because those types (strings, lists, etc.) are far too simple and limited. Consider, for instance, the task of representing a protein 3D structure, where 'representing' means storing all the information that one may wish to access and manipulate: AA sequence (residue types and numbers), the atoms comprising each residue, the spatial coordinates of each atom, whether a cysteine residue is disulfide-bonded or not, the protein's function, the year the protein was discovered, a list of orthologs of known structure, and so on. What data structure might be capable of most naturally representing such an entity? A simple (generic) Python tuple or list is clearly insufficient.

For this problem, one could try to represent the protein as a single tuple, where the first element is a list of the sequence of residues, the second element is a string describing the protein's function, the third element lists orthologs, etc. Somewhere within this top-level list, the coordinates of the C$_\alpha$ atom of Alanine-42 might be represented as `[x, y, z]`, which is a simple list of length three. (The list is 'simple' in the sense that its rank is one; the *rank* of a tuple or list is, loosely, the number of dimensions spanned by its rows, and in this case we have but one row.) In other words, our overall data-representation problem can be hierarchically decomposed into simpler sub-problems that **are** amenable to representation via Python's built-in types. While valid, such a data structure will be difficult to use: The programmer will have to recall multiple arbitrary numbers (list and sub-list indices) in order to access anything, and extensions to this approach will only make it clumsier. Additionally, there are many functions that are meaningful only in the context of proteins, not all tuples. For example, we may need to compute the solvent-accessible surface areas of all residues in all β-strands for a list of proteins, but this operation would be nonsensical for a list of Supreme Court cases. Conversely, not all tuple methods would be relevant to this protein data structure, yet a function to find cases that reached a 5-4 decision along party lines would accept the protein as an argument. In other words, the tuple mentioned above has no clean way to make the necessary associations. It's just a tuple.

## OOP Terminology

This protein representation problem is elegantly solved via the OOP concepts of classes, objects, and methods. Briefly, an *object* is an instance of a data structure that contains members and methods. *Members* are data of potentially any type, including other objects. Unlike lists and tuples, where the elements are indexed by numbers starting from zero, the members of an object are given names, such as `yearDiscovered`. *Methods* are functions that (typically) make use of the members of the object. Methods perform operations that are related to the data in the object's members. Objects are constructed from *class* definitions, which are blocks that define what most of the methods will be for an object. The next section's examples (§4.4) will help clarify this terminology. (Finally, note that some languages require that all methods and members be specified in the class declaration, but Python allows *duck punching*, or adding members after declaring a class. Adding methods later is possible too, but uncommon. Some built-in types, such as `int`, do not support duck punching.)

During execution of an actual program, a specific object is created by calling the name of the class, as one would do for a function. The interpreter will set aside some memory for the object's methods and members, and then call a method named `__init__`, which initializes the object for use.

Classes can be created from previously defined classes. In such cases, all properties of the parent class are said to be *inherited* by the child class. The child class is termed a *derived class*, while the parent is described as a *base class*. For instance, a user-defined `Biopolymer` class may have derived classes named `Protein` and `NucleicAcid`, and may itself be derived from a more general `Molecule` base class. Class names often begin with a capital letter, while object names (i.e., variables) often start with a lowercase letter. Within a class definition, a leading underscore denotes member names that will be protected. Working examples and annotated descriptions of these concepts can be found, in the context of protein structural analysis, in ref [85].

The OOP paradigm suffuses the Python language: Every value is an object. For example, the statement `foo='bar'` instantiates a new object (of type `str`) and binds the name `foo` to that object. All built-in string methods will be exposed for that object (e.g., `print(foo.upper())` returns `'BAR'`). Python's built-in `dir()` function can be used to list all attributes and methods of an object, so `dir(foo)` will list all available attributes and valid methods on the variable `foo`. The statement `dir(1)` will show all the methods and members of an `int` (there are many!). This example also illustrates the conventional OOP dot-notation, `object.attribute`, which is used to access an object's members, and to invoke its methods (Fig. 1, left). For instance, `protein1.residues[2].CA.x` might give the $x$-coordinate of the C$_\alpha$ atom of the third residue in `protein1` as a floating-point number, and `protein1.residues[5].ssbond(protein2.residues[6])` might be used to define a disulfide bond (the `ssbond()` method) between residue-6 of `protein1` and residue-7 of `protein2`. In this example, the `residues` member is a list or tuple of objects, and an item is retrieved from the collection using an index in brackets.

## Benefits of OOP

By effectively compartmentalizing the programming logic and implicitly requiring a disciplined approach to data structures, the OOP paradigm offers several benefits. Chief among these are (i) clean data/code separation and bundling (i.e., modularization), (ii) code reusability, (iii) greater extensibility (derived classes can be created as needs become more specialized), and (iv) encapsulation into classes/objects provides a clearer interface for other programmers and users. Indeed, a generally good practice is to discourage end-users from

PLOS 25/45

directly accessing and modifying all of the members of an object. Instead, one can expose a limited and clean interface to the user, while the back-end functionality (which defines the class) remains safely under the control of the class' author. As an example, custom *getter* and *setter* methods can be specified in the class definition itself, and these methods can be called in another user's code in order to enable the safe and controlled access/modification of the object's members. A setter can 'sanity-check' its input to verify that the values do not send the object into a nonsensical or broken state; e.g., specifying the string `"ham"` as the $x$-coordinate of an atom could be caught before program execution continues with a corrupted object. By forcing alterations and other interactions with an object to occur via a limited number of well-defined getters/setters, one can ensure that the integrity of the object's data structure is preserved for downstream usage.

The OOP paradigm also solves the aforementioned problem wherein a protein implemented as a tuple had no good way to be associated with the appropriate functions—we could call Python's built-in `max()` on a protein, which would be meaningless, or we could try to compute the isoelectric point of an arbitrary list (of Supreme Court cases), which would be similarly nonsensical. Using classes sidesteps these problems. If our `Protein` class does not define a `max()` method, then no attempt can be made to calculate its maximum. If it does define an `isoelectricPoint()` method, then that method can be applied only to an object of type `Protein`. For users/programmers, this is invaluable: If a class from a library has a particular method, one can be assured that that method will work with objects of that class.

### OOP in Practice: Some Examples

A classic example of a data structure that is naturally implemented via OOP is the creation of a `Human` class. Each `Human` object can be fully characterized by her respective properties (members such as height, weight, etc.) and functionality (methods such as breathing, eating, speaking, etc.). A specific human being, e.g. `guidoVanRossum`, is an instance of the `Human` class; this class may, itself, be a subclass of a `Hominidae` base class. The following code illustrates how one might define a `Human` class, including some functionality to age the `Human` and to set/get various members (descriptors such as height, age, etc.):

```python
from random import randint

class Human():
  _age = 0
  _height = 0
  _sex = ""

  def __init__(self, theSex):
    self.setSex(theSex)
  def haveBirthday(self):
    self._age += 1
    if (self._age < 21):
      self._height += randint(0.15)
    if (self._age > 60):
      self._height -= randint(0.05)
  def setAge(self,theAge):
    assert(theAge >= 0)
    self._age = theAge
  def setHeight(self,theHeight):
    self._height = theHeight
  def setSex(self,theSex):
    assert(theSex == "male" or theSex == "female") # Validate input.
    self._sex = theSex
  def getAge(self):
```

```python
25      return self._age
26   def getHeight(self):
27      return self._height
28   def getSex(self):
29      return self._sex
30 # now use the class by instantiating an object and manipulating it:
31 guido=Human("male")      # This will call __init__ and set guido's sex to "male".
32 print(guido.getAge())    # see the default value of age.
   0
33 guido.setAge(21)         # set guido's age.
34 guido.haveBirthday()     # it's guido's birthday.
35 print(guido.getAge())    # verify that age has advanced by one unit.
   22
```

Note the usage of `self` as the first argument in each method defined in the above code. The `self` keyword is necessary because when a method is invoked it must know **which** object to use. That is, an object instantiated from a class requires that methods on that object have some way to reference that particular instance of the class, versus other potential instances of that object. The `self` keyword provides such a 'hook' to reference the specific object for which a method is called. Every method invocation for a given object, including even the initializer called __init__, must pass it**self** (the current instance) as the first argument to the method; this subtlety is further described at [86] and [87]. A practical way to view the effect of `self` is that any occurrence of `objName.methodName(arg1,arg2)` effectively becomes `methodName(objName,arg1,arg2)`. This is one key deviation from the behavior of top-level functions, which exist outside of any class. When defining methods, usage of `self` provides an explicit way for the object itself to be provided as an argument (self-reference), and its disciplined usage will help minimize confusion about expected arguments.

To illustrate how objects may interact with one another, consider a class to represent a chemical's atom:

```python
1 class Atom:
2   (x,y,z) = (0,0,0)   # compact way to set all three vars simultaneously (rather
3                       # than on three separate lines, as x=0, y=0, z=0)
4   name = "X"
5   def distanceFrom(self, other):
6     return ((self.x - other.x)**2 +
7             (self.y - other.y)**2 +
8             (self.z - other.z)**2)**0.5
9   def setName(self, newName):
10    assert(newName in ("C", "H", "O", "N", "S"))
11    self.name = newName
12  def moveTo(self, newX, newY, newZ):
13    self.x = newX
14    self.y = newY
15    self.z = newZ
```

Then, we can use this `Atom` class in constructing another class to represent molecules:

```python
1 class Molecule:
2   atoms = []
3   bonds = []
4   def addAtom(self, newAtom):    # newAtom will be an Atom object
5     self.atoms.append(newAtom)
6   def makeBond(self, a1, a2):
7     self.bonds.append((a1, a2))
8   def avgBondLength(self):
9     totLength = 0
```

```
10      for bondIndex in range(0,len(self.bonds)):
11          firstAtom = self.bonds[bondIndex][0]
12          secondAtom = self.bonds[bondIndex][1]
13          totLength += firstAtom.distanceFrom(secondAtom)
14      return totLength / len(self.bonds)
```

And, finally, the following code illustrates the construction of a diatomic molecule:

```
1 a1=Atom(); a2=Atom()    # instantiate   two atoms
2 a2.moveTo(1,1,0)        # move the second atom to (1,1,0)
3 m=Molecule()            # instantiate   a new molecule
4 m.addAtom(a1)           # ... and populate it with
5 m.addAtom(a2)           #      these two atoms
6 m.makeBond(a1,a2)       # define a bond between the two atoms
7 print (m.avgBondLength())
  1.4142135623730951
```

If the above code is run, for example, in an interactive Python session, then note that the aforementioned `dir()` function is an especially useful built-in tool for querying the properties of new classes and objects. For instance, issuing the statement `dir(Molecule)` will return detailed information about the `Molecule` class (including its available methods).

**Exercise 9**: Amino acids can be effectively represented via OOP because each AA has a well-defined chemical composition: a specific number of atoms of various element types (carbon, nitrogen, etc.) and a covalent bond connectivity that adheres to a specific pattern. For these reasons, the prototype of an L-amino acid can be unambiguously defined by the SMILES [88] string 'N[C@@H](R)C(=O)O', where 'R' denotes the side-chain and '@@' indicates the L enantiomer. In addition to chemical structure, each AA also features specific physicochemical properties (molar mass, isoelectric point, optical activity/specific rotation, etc.). In this exercise, create an AA class and use it to define any two of the twenty standard AAs, in terms of their chemical composition and unique physical properties. To extend this exercise, consider expanding your AA class to include additional class members (e.g., the frequency of occurrence of that AA type) and methods (e.g., the possibility of applying post-translational modifications). To see the utility of this exercise in a broader OOP schema, see the discussion of the hierarchical **S**tructure ⊃ **M**odel ⊃ **C**hain ⊃ **R**esidue ⊃ **A**tom (SMCRA) design used in ref [85] to create classes that can represent entire protein assemblies.

## File Management and I/O

Scientific data are typically acquired, processed, stored, exchanged, and archived as computer files. As a means of input/output (I/O) communication, Python provides tools for reading, writing and otherwise manipulating files in various formats. Supplemental Chapter 11 focuses on file I/O in Python. Most simply, the Python interpreter allows command-line input and basic data output via the `print()` function. For real-time interaction with Python, the free IPython [89] system offers a shell that is both easy to use and uniquely powerful (e.g., it features tab completion and command history scrolling); see the S2 Text, §3 for more on interacting with Python. A more general approach to I/O, and a more robust (persistent) approach to data archival and exchange, is to use files for reading, writing, and processing data. Python handles file I/O via the creation of `file` objects, which are instantiated by calling the `open` function with the filename and access mode as its two arguments. The syntax is illustrated by `fileObject = open("myName.pdb", mode='r')`, which creates a new file object from a file named `"myName.pdb"`. This file will be only readable because the `'r'` mode is specified; other valid modes include `'w'` to allow writing and `'a'` for appending. Depending on which mode is specified, different methods of the file object will be exposed for use. Table 4 describes mode types and the various methods of a `file` object.

**Table 4 Python's File-access Modes**

| I/O Mode | Syntax | Behavior |
|---|---|---|
| Read | 'r' | Opens the contents of a file for reading into the file interface, allowing for lines to be read-in successively. |
| Write | 'w' | Creates a file with the specified name and allows for text to be written to the file; note that specifying a pre-existing filename will overwrite the existing file. |
| Append | 'a' | Opens an existing file and allows for text to be written to it, starting at the conclusion of the original file contents. |
| Read and Write | 'r+' | Opens a file such that its contents can be both read-in and written-to, thus offering great versatility. |

Python's available file-access modes are summarized here.

The following example opens a file named `myDataFile.txt` and reads the lines, *en masse*, into a list named `listOfLines`. (In this example, the variable `readFile` is also known as a 'file handle', as it references the file object.) As for all lists, this object is iterable and can be looped over in order to process the data.

```
readFile = open("myDataFile.txt", mode='r')
listOfLines = readFile.readlines()
# Process the lines. Simply dump the contents to the console:
for l in listOfLines:
   print(l)
 (The lines in the file will be printed)
readFile.close()
```

Data can be extracted and processed via subsequent string operations on the list of lines drawn from the file. In fact, many data-analysis workflows commit much effort to the pre-processing of raw data and standardization of formats, simply to enable data structures to be cleanly populated. For many common input formats such as .csv (comma-separated values) and .xls (Microsoft Excel), packages such as `pandas` [90] simplify the process of reading in complex file formats and organizing the input as flexible data structures. For more specialized file formats, much of this 'data wrangling' stems from the different degrees of standards-compliance of various data sources, as well as the immense heterogeneity of modern collections of datasets (sequences, 3D structures, microarray data, network graphs, etc.). A common example of the need to read and extract information is provided by the PDB file format [22], which is a container for macromolecular structural data. In addition to its basic information content—lists of atoms and their 3D coordinates—the standard PDB file format also includes a host of *metadata* (loosely, data that describe other (lower-level) data, for instance in terms of syntax and schemas), such as the biopolymer sequence, protein superfamily, quaternary structures, chemical moieties that may be present, X-ray or NMR refinement details, and so on. Indeed, processing and analyzing the rich data available in a PDB file motivates this primer's Final Project (§11). For now, this brief example demonstrates how to use Python's I/O methods to count the number of HETATM records in a PDB file:

```
fp = open('1I8F.pdb', mode='r')
numHetatm = 0
for line in fp.readlines():
   if(len(line) > 6):
     if(line[0:6] == "HETATM"):
        numHetatm += 1
fp.close()
print(numHetatm)
 160
```

Such HETATM, or heteroatom, lines in a PDB file correspond to water, ions, small-molecule ligands, and other non-biopolymer components of a structure; for example, glycerol HETATM lines are often found in cryo-crystallographic structures, where glycerol was added to crystals as a cryo-protectant.

**Exercise 10**: The standard FASTA file-format, used to represent protein and nucleic acid sequences, consists of two parts: (i) The first line is a description of the biomolecule, starting with a greater-than sign (>) in the first column; this sign is immediately followed by a non-whitespace character and any arbitrary text that describes the sequence name and other information (e.g., database accession identifiers). (ii) The subsequent lines specify the biomolecular sequence as single-letter codes, with no blank lines allowed. A protein example follows:

```
>tr|Q8ZYG5|Q8ZYG5_PYRAE (Sm-like) OS=Pyrobaculum aerophilum GN=PAE0790
MASDISKCFATLGATLQDSIGKQVLVKLRDSHEIRGILRSFDQHVNLLLEDAEEIIDGNV
YKRGTMVVRGENVLFISPVP
```

Begin this exercise by choosing a FASTA protein sequence with more than 3000 AA residues. Then, write Python code to read in the sequence from the FASTA file and: (i) determine the relative frequencies of AAs that follow proline in the sequence; (ii) compare the distribution of AAs that follow proline to the distribution of AAs in the entire protein; and (iii) write these results to a human-readable file.

## Regular Expressions for String Manipulations

The *regular expression* (*regex*) is an extensible tool for pattern matching in strings. They are discussed at length in Supplemental Chapter 17. Regexes entered the world of practical programming in the late 1960s at Bell Labs and, like many tools of that era, they are powerful, flexible, and terse constructs. Fundamentally, a regex specifies a set of strings. The simplest type of regex is a simple string with no special characters (*metacharacters*). Such a regex will match itself: `Biology` would match 'Biology' or 'Biologys', but not 'biology', 'Biochem', or anything else that does not start with 'Biology' (note the case sensitivity).

In Python, a regex `match`es a string if the string starts with that regex. Python also provides a `search` function to locate a regex anywhere within a string. Returning to the notion that a regex "specifies a set of strings", given some text the `match`es to a regex will be all strings that **start** with the regex, while the `search` hits will be all strings that **contain** the regex. For clarity, we will say that a regex `find`s a string if the string is completely described by the regex, with no trailing characters. (There is no `find` in Python but, for purposes of description here, it is useful to have a term to refer to a `match` without trailing characters.)

Locating strings and parsing text files is a ubiquitous task in the biosciences, e.g. identifying a stop codon in a nucleic acid FASTA file or finding error messages in an instrument's log files. Yet regexes offer even greater functionality than may be initially apparent from these examples, as described below. First, we note that the following metacharacters are special in regexes: `$^.*+?{}[]()|\`, and in most cases they do not `find` themselves.

The `^` and `$` metacharacters (known as *anchors*) are straightforward, as they `find` the start and end of a line, respectively. While `match` looks for lines beginning with the specified regex, adding a `$` to the end of the regex pattern will ensure that any matching line ends at the end of the regex. (This is why there is no `find` function in Python: it is easily achieved by adding a `$` to a regex used in `match`.) For example, to `find` lines in a log file that state 'Run complete', but not 'Run completes in 5 minutes', the regex `Run complete$` would `match` the desired target lines.

A `.` (a period) `find`s literally any character. For example, if a protein kinase has a consensus motif of 'AXRSXRSXRSP', where X is any AA, then the regex `A.RS.RS.RSP` would succeed in `search`ing for substrates.

The metacharacters `*`, `+`, `{}`, and `?` are special *quantifier* operators, used to specify repetition of a character, character class, or higher-order unit within a regex (described below).

A `*` after a character (or group of characters) `finds` that character zero or more times. Returning to the notion of a consensus motif, a protein that recognizes RNA which contains the dinucleotide 'UG' followed by any number of 'A's would find its binding partners by `searching` for the regex `UGA*`. One can comb through RNA-seq reads to find sequences that are 3'-polyadenylated by `searching` for `AAAAAA*$`. This would `find` exactly five 'A's, followed by zero or more 'A's, followed by the end of the line. The `+` metacharacter is akin to `*`, except that it `finds` one or more of the preceding character. A `?` `finds` the preceding character zero or one time. Most generally, the `{m,n}` syntax `finds` the preceding character (possibly from a character class) between $m$ and $n$ times, inclusive. Thus, `x{3}` `finds` the character 'x' if repeated exactly three times, `A{5,18}` `finds` the character 'A' repeated five to eighteen times, and `P{2,}` `finds` runs of two or more 'P' characters.

Combining the above concepts, we can `search` for protein sequences that begin with a His6×-tag ('HHHHHH'), followed by at most five residues, then a TEV protease cleavage site ('ENLYFQ'), followed immediately by a 73-residue polypeptide that ends with 'IIDGNV'. The regex to `search` for this sequence would be `H{6}.{0,5}ENLYFQ.{67}IIDGNV`.

Characters enclosed in square brackets, `[]`, specify a *character class*. This functionality allows a regex to `find` any of a set of characters. For example, `AAG[TC]G` would `find` 'AAG**T**G' or 'AAG**C**G', where the variable char from the character class is bolded. A range of characters can be provided by separating them with a hyphen, `-`. So, for instance, `[A-Z][a-z]*` would `find` a word that starts with a capital letter. Multiple ranges can be specified, and `[1-9][A-Za-z0-9]{3}.pdb` would `find` PDB files in some search directory. (Note that the `.` in that regex will `find` any character, so '1l8Fnpdb' would be matched, even though we might intend for only '1l8F.pdb' to be found. This could be corrected by escaping the `.` with a backslash, as discussed below.) The `^` metacharacter can be used to negate a character class: `[^0-9]` would `find` any non-numeric character.

The backslash metacharacter, `\`, is used to suppress, or *escape*, the meaning of the immediately following character; for this reason, `\` is known as an *escape character*. For example, consider the task of finding prices exceeding $1000 in a segment of text. A regex might be `$0*[1-9][0-9]{3,}.[0-9]{2}`. This monstrous regex should `find` a dollar sign, any number of zeros, one non-zero number, at least three numbers, a period, and two numbers. Thus, '$01325.25' would be found, but not '$00125.67'. (The requirement of a non-zero number followed by three numbers is not met in this case.) But, there is a problem here: The `$` metacharacter anchors the end of a line, and because no text can appear after the end of a line this regex will never `match` any text. Furthermore, the `.` is meant to `find` a literal period (the decimal point), but in a regex it is a wildcard that `finds` any character. The `\` metacharacter can be used to solve these problems: It notifies the regex engine to treat the subsequent character as a literal. Thus, a correct regex for prices over $1000 would be `\$0*[1-9][0-9]{3,}\.[0-9]{2}`. To `find` a literal '\', use `\\`. (The `\` metacharacter often appears in I/O processing as a way to escape quotation marks; for instance, the statement `print("foo")` will output `foo`, whereas `print("\"foo\"")` will print `"foo"`.)

Python and many other languages include a `\` before certain (non-reserved) characters, as a convenient built-in feature for commonly-used character classes. In particular, `\d` `finds` any digit, `\s` `finds` whitespace, `\S` `finds` non-whitespace, and `\w` `finds` any alphanumeric character or underscore (i.e., the class `[a-zA-Z0-9_]`), such as typically occurs in ordinary English words. These built-in features can be used to more compactly express the price regex, including the possibility of whitespace between the '$' sign and the first digit: `\$\s*0*[1-9]\d{3,}\.\d{2}`.

The `|` metacharacter is the logical 'or' operator (also known as the *alternation* operator). The regex `abc|xyz` will find either 'abc' or 'xyz'. Initially, the behavior of `|` can be deceptive: `£|€|$ [0-9]*` is not equivalent to `[£€$] [0-9]*`, as the former will `find` a lone pound symbol, a lone Euro symbol, or a dollar sign followed by a number. As an

example, to `match` the SEQRES and ATOM records in a PDB file, `ATOM.*|SEQRES.*` would work.

The final metacharacters that we will explore are matched parentheses, `()`, which `find` *character groups*. While `x[abc]y` will `find` 'xay', 'xby', or 'xcy', the regex `x(abc)y` `match`es only those strings starting with 'xabcy'—i.e., it is equivalent to `xabcy`. The utility of groups stems from the ability to use them as units of repetition. For example, to see if a sequence is delimited by a start and stop codon, and therefore is a potential ORF, we could use `AUG.*U(AA|AG|GA)`; this regex will `search` for 'UAA', 'UAG', or 'UGA' at the end of the sequence. (Note that parentheses delimit the `|`.) Note that this regex does not check that the start and stop codon are in the same frame, since the characters that `find` captures by the `.*` may not be a multiple of three. To address this, the regex could be changed to `AUG(...)*U(AA|AG|GA)`. Another feature of groups is the ability to refer to previous occurrences of a group within the regex (a *backreference*), enabling even more versatile pattern matching. To explore groups and other powerful features of regexes, readers can consult thorough texts [91] and numerous online resources (e.g., [92, 93]).

Beyond the central role of the regex in analyzing biological sequences, parsing datasets, etc., note that any effort spent learning Python regexes is highly transferable. In terms of general syntactic forms and functionality, regexes behave roughly similarly in Python and in many other mainstream languages (e.g., Perl, R), as well as in the shell scripts and command-line utilities (e.g., `grep`) found in the Unix family of operating systems (including all Linux distributions and Apple's OS X).

**Exercise 11**: Many human hereditary neurodegenerative disorders, such as Huntington's disease (HD), are linked to anomalous expansions in the number of trinucleotide repeats in particular genes [94]. In HD, the pathological severity correlates with the number of $(CAG)_n$ repeats in exon-1 of the gene (*htt*) encoding the protein (huntingtin): More repeats means an earlier age of onset and a more rapid disease progression. The CAG codon specifies glutamine, and HD belongs to a broad class of polyglutamine (polyQ) diseases. Healthy (wild-type) variants of this gene feature $n \approx$ 6-35 tandem repeats, whereas $n > 35$ virtually assures the disease. For this exercise, write a Python regex that will locate any consecutive runs of $(CAG)_{n>10}$ in an input DNA sequence. Because the codon CAA also encodes Q and has been found in long runs of CAGs, your regex should also allow interspersed CAAs. To extend this exercise, write code that uses your regex to count the number of CAG repeats (allow CAA too), and apply it to a publically-available genome sequence of your choosing (e.g., the NCBI GI code 588282786:1-585 is exon-1 from a human's *htt* gene [accessible at `http://1.usa.gov/1NjrDNJ`]).

# An Advanced Vignette: Creating Graphical User Interfaces with Tkinter

Thus far, this primer has centered on Python programming as a tool for interacting with data and processing information. To illustrate an advanced topic, this section shifts the focus towards approaches for creating software that relies on user interaction, via the development of a graphical user interface (GUI; pronounced 'gooey'). Text-based interfaces (e.g., the Python shell) have several distinct advantages over purely graphical interfaces, but such interfaces can be intimidating to the uninitiated. For this reason, many general users will prefer GUI-based software that permits options to be configured via graphical check boxes, radio buttons, pull-down menus and the like, versus text-based software that requires typing commands and editing configuration files. In Python, the `tkinter` package (pronounced 'T-K-inter') provides a set of tools to create GUIs. (Python 2.x calls this package `Tkinter`, with a capital `T`; here, we use the Python 3.x notation.)

Tkinter programming has its own specialized vocabulary. *Widgets* are objects, such as text boxes, buttons and frames, that comprise the user interface. The *root window* is the widget that contains all other widgets. The root window is responsible for monitoring user interactions and informing the contained widgets to respond when the user triggers an interaction with them (called an *event*). A *frame* is a widget that contains other widgets. Frames are used to group related widgets together, both in the code and on-screen. A *geometry manager* is a system that places widgets in a frame according to some style determined by the programmer. For example, the `grid` geometry manager arranges widgets on a grid, while the `pack` geometry manager places widgets in unoccupied space. Geometry managers are discussed at length in Supplemental Chapter 18, which shows how intricate layouts can be generated.

The basic style of GUI programming fundamentally differs from the material presented thus far. The reason for this is that the programmer cannot predict what actions a user might perform, and, more importantly, in what order those actions will occur. As a result, GUI programming consists of placing a set of widgets on the screen and providing instructions that the widgets execute when a user interaction triggers an event. (Similar techniques are used, for instance, to create web interfaces and widgets in languages such as JavaScript.) Supplemental Chapter 19 describes available techniques for providing functionality to widgets. Once the widgets are configured, the root window then awaits user input. A simple example follows:

```
from tkinter import Tk, Button
def buttonWindow():
  window = Tk()
  def onClick():
    print("Button clicked")
  btn = Button(window, text="Sample Button", command=onClick)
  btn.pack()
  window.mainloop()
```

To spawn the Tk window, enter the above code in a Python shell and then issue the statement `buttonWindow()`. Then, press the 'Sample Button' while viewing the output on the console. The first line in the above code imports the `Tk` and `Button` classes. `Tk` will form the root window, and `Button` will create a button widget. Inside the function, line 3 creates the root window. Lines 4 and 5 define a function that the button will call when the user interacts with it. Line 6 creates the button. The first argument to the `Button` constructor is the widget that will contain the button, and in this case the button is placed directly in the root window. The `text` argument specifies the text to be displayed on the button widget. The `command` argument attaches the function named `onClick` to the button. When the user presses the button, the root window will instruct the button widget to call this function. Line 7 uses the `pack` geometry manager to place the button in the root window. Finally, line 8 instructs the root window to enter `mainloop`, and the root window is said to *listen* for user input until the window is closed.

Graphical widgets, such as text entry fields and check-boxes, receive data from the user, and must communicate that data within the program. To provide a conduit for this information, the programmer must provide a variable to the widget. When the value in the widget changes, the widget will update the variable and the program can read it. Conversely, when the program should change the data in a widget (e.g., to indicate the status of a real-time calculation), the programmer sets the value of the variable and the variable updates the value displayed on the widget. This roundabout tack is a result of differences in the architecture of Python and Tkinter—an integer in Python is represented differently than an integer in Tkinter, so reading the widget's value directly would result in a nonsensical Python value. These variables are discussed in Supplemental Chapter 19.

From a software engineering perspective, a drawback to graphical interfaces is that multiple GUIs cannot be readily composed into new programs. For instance, a GUI to display how a particular restriction enzyme will cleave a DNA sequence will not be practically useful in predicting the products of digesting thousands of sequences with the enzyme, even though some core component of the program (the key, non-GUI program logic) would be useful in automating that task. For this reason, GUI applications should be written in as modular a style as possible—one should be able to extract the useful functionality without interacting with the GUI-specific code. In the restriction enzyme example, an optimal solution would be to write the code that computes cleavage sites as a separate module, and then have the GUI code interact with the components of that module.

## Python in General-purpose Scientific Computing: Numerical Efficiency, Libraries

In pursuing biological research, the computational tasks that arise will likely resemble problems that have already been solved, problems for which software libraries already exist. This occurs largely because of the interdisciplinary nature of biological research, wherein relatively well-established formalisms and algorithms from physics, computer science, and mathematics are applied to biological systems. For instance, (i) the simulated annealing method was developed as a physically-inspired approach to combinatorial optimization, and soon thereafter became a cornerstone in the refinement of biomolecular structures determined by NMR spectroscopy or X-ray crystallography [95]; (ii) dynamic programming was devised as an optimization approach in operations research, before becoming ubiquitous in sequence alignment algorithms and other areas of bioinformatics; and (iii) the Monte Carlo method, invented as a sampling approach in physics, underlies the algorithms used in problems ranging from protein structure prediction to phylogenetic tree estimation.

Each computational approach listed above can be implemented in Python. The language is well-suited to rapidly develop and prototype any algorithm, be it intended for a relatively lightweight problem or one that is more computationally intensive (see [96] for a text on general-purpose scientific computing in Python). When considering Python and other possible languages for a project, software development time must be balanced against a program's execution time. These two factors are generally countervailing because of the inherent performance trade-offs between codes that are written in interpreted (high-level) versus compiled (lower-level) languages; ultimately, the computational demands of a problem will help guide the choice of language. In practice, the feasibility of a pure Python versus non-Python approach can be practically explored via numerical benchmarking. While Python enables rapid development, and is of sufficient computational speed for many bioinformatics problems, its performance simply cannot match the compiled languages that are traditionally used for high-performance computing applications (e.g., many MD integrators are written in C or Fortran). Nevertheless, Python codes are available for molecular simulations, parallel execution, and so on. Python's popularity and utility in the biosciences can be attributed to its ease of use (expressiveness), its adequate numerical efficiency for many bioinformatics calculations, and the availability of numerous libraries that can be readily integrated into one's Python code (and, conversely, one's Python code can 'hook' into the APIs of larger software tools, such as PyMOL). Finally, note that rapidly-developed Python software can be integrated with numerically efficient, high-performance code written in a low-level languages such as C, in an approach known as 'mixed-language programming' [49].

Many third-party Python libraries are now well-established. In general, these mature projects are (i) well-documented, (ii) freely available as stable (production) releases, (iii) undergoing continual development to add new features, and (iv) characterized by large user-bases and active communities (mailing lists, etc.). A useful collection of such tools can be found at the SciPy resource [97, 98], which is a platform for the maintenance and distribution of several popular packages: (i) NumPy, which is invaluable for matrix-related calculations [99]; (ii) SciPy, which provides routines from linear algebra, signal processing, statistics, and a wealth of other numerical tools; (iii) pandas, which facilitates data import, management, and organization [90]; and (iv) matplotlib, a premiere codebase for plotting and general-purpose visualization [62]. The package scikit-learn extends SciPy with machine learning and statistical analysis functionalities [100]. Other statistical tools are available in the statistics standard library, in SciPy [97, 98], and in NumPy [99]; finally, many more-specialized packages also exist, such as pyBrain [78] and DEAP [101]. Properly interacting with Python modules, such as those mentioned above, is detailed in Supplemental Chapter 4 (S1 Text).

Many additional libraries can be found at the official Python Package Index (PyPI; [102]), as well as myriad packages from unofficial third-party repositories. The aforementioned (§1.4) BioPython project offers an integrated suite of tools for sequence- and structure-based bioinformatics, as well as phylogenetics, machine learning, and other feature sets. We survey the computational biology software landscape in the S2 Text (§2), including tools for structural bioinformatics, phylogenetics, omics-scale data-processing pipelines, and workflow management systems. Finally, note that Python code can be interfaced with other languages. For instance, current support is provided for low-level integration of Python and R [103, 104], as well as C-extensions in Python (Cython; [105, 106]). Such cross-language interfaces extend Python's versatility and flexibility for computational problems at the intersection of multiple scientific domains, as often occurs in the biosciences.

## Python and Software Licensing

Any discussion of libraries, modules, and extensions merits a brief note on the important role of licenses in scientific software development. As evidenced by the widespread utility of existing software libraries in modern research communities, the development work done by one scientist will almost certainly aid the research pursuits of others—either near-term or long-term, in subfields that might be near to one's own or perhaps more distant (and unforeseen). *Free software* licenses promote the unfettered advance of scientific research by encouraging the open exchange, transparency, communicability, and reproducibility of research projects. To qualify as free software, a program must allow the user to view and change the source code (for any purpose), distribute the code to others, and distribute modified versions of the code to others. The Open Source Initiative provides alphabetized and categorized lists of licenses that comply, to various degrees, with the open-source definition [107]. As an example, the Python interpreter, itself, is under a free license. Software licensing is a major topic unto itself, and helpful primers are available on technical [38] and strategic [37, 108] considerations in adopting one licensing scheme versus another. All of the content (code and comments) that is provided as Supplemental Chapters with this text is licensed under the GNU Affero General Public License (AGPL) version 3, which permits anyone to examine, edit, and distribute the source so long as any works using it are released under the same license.

## Managing Large Projects: Version Control Systems

As a project grows, it becomes increasingly difficult—yet increasingly important—to be able to track changes in source code. A *version control system* (VCS) tracks changes to documents and facilitates the sharing of code among multiple individuals. In a *distributed* (as opposed to centralized) VCS, each developer has his own complete copy of the project, locally stored. Such a VCS supports the 'committing', 'pulling', 'branching', and 'merging' of code. After making a change, the programmer *commits* the change to the VCS. The VCS stores a snapshot of the project, preserving the development history. If it is later discovered that a particular commit introduced a bug, one can easily roll-back the offending commit. Other developers who are working on the same project can *pull* from the author of the change (the most recent version, or any earlier snapshot). The VCS will incorporate the changes made by the author into the puller's copy of the project. If a new feature will make the code temporarily unusable (until the feature is completely implemented), then that feature should be developed in a separate *branch*. Developers can switch between branches at will, and a commit made to one branch will not affect other branches. The *master branch* will still contain a working version of the program, and developers can still commit non-breaking changes to the master branch. Once the new feature is complete, the branches can be *merged* together. In a distributed VCS, each developer is, conceptually, a branch. When one developer pulls from others, this is equivalent to merging a branch from each developer. Git, Mercurial, and Darcs are common distributed VCSs. In contrast, in a *centralized* VCS all commits are tracked in one central place (for both distributed and centralized VCSs, this 'place' is often a repository hosted in the cloud). When a developer makes a commit, it is pushed to every other developer (who is on the same branch). The essential behaviors—committing, branching, merging—are otherwise the same as for a distributed VCS. Examples of popular centralized VCSs include the Concurrent Versioning System (CVS) and Subversion.

While VCSs are mainly designed to work with source code, they are not limited to this type of file. A VCS is useful in many situations where multiple people are collaborating on a single project, as it simplifies the task of combining, tracking, and otherwise reconciling the contributions of each person. In fact, this very document was developed using LaTeX and the Git VCS, enabling each author to work on the text in parallel. A helpful guide to Git and GitHub (a popular Git repository hosting service) was very recently published [109]; in addition to a general introduction to VCSs, that guide offers extensive practical advice, such as what types of data/files are more or less ideal for version controlling.

## Final Project: A Structural Bioinformatics Problem

Fluency in a programming language is developed actively, not passively. The exercises provided in this text have aimed to develop the reader's command of basic features of the Python language. Most of these topics are covered more deeply in the Supplemental Chapters, which also include some advanced features of the language that lie beyond the scope of the main body of this primer. As a final exercise, a cumulative project is presented below. This project addresses a substantive scientific question, and its successful completion requires one to apply and integrate the skills from the foregoing exercises. Note that a project such as this—and really any project involving more than a few dozen lines of code—will benefit greatly from an initial planning phase. In this initial stage of software design, one should consider the basic functions, classes, algorithms, control flow and overall code structure.

**Exercise 12** (cumulative project): First, obtain a set of several hundred protein structures from the PDB, as plaintext .pdb files (the exact number of entries is immaterial). Then, from this pool of data, determine the relative frequencies of the constituent amino acids for each protein secondary structural class; use only the three descriptors 'helix', 'sheet', and, for any AA not within a helix or sheet, 'irregular'. (Hint: In considering file parsing and potential data structures, search online for the PDB's file-format specifications.) Output your statistical data to a human-readable file format (e.g., comma-separated values, .csv) such that the results can be opened in a statistical or graphical software package for further processing and analysis. As a bonus exercise, use Python's matplotlib package to visualize the findings of your structural bioinformatics analysis.

# Conclusion

Data and algorithms are two pillars of modern biosciences. Data are acquired, filtered, and otherwise manipulated in preparation for further processing, and algorithms are applied in analyzing datasets so as to obtain results. In this way, computational workflows transform primary data into results that can, over time, become formulated into general principles and new knowledge. In the biosciences, modern scientific datasets are voluminous and heterogeneous. Thus, in developing and applying computational tools for data analysis, the two central goals are **scalability**, for handling the data-volume problem, and **robust abstractions**, for handling data heterogeneity and integration. These two challenges are particularly vexing in biology, and are exacerbated by the traditional lack of training in computational and quantitative methods in many biosciences curricula. Motivated by these factors, this primer has sought to introduce general principles of computer programming, at both basic and intermediate levels. The Python language was adopted for this purpose because of its broad prevalence and deep utility in the biosciences.

# Supporting Information

### S1 Text (Python Chapters)

This suite of 19 Supplemental Chapters covers the essentials of programming. The Chapters are written in Python, and guide the reader through the core concepts of programming, via numerous examples and explanations. The most recent versions of all materials are maintained at `http://p4b.muralab.org`. For purposes of self-study, solutions to the in-text exercises are also included.

### S2 Text

The supplemental text contains sections on: *(i)* Python as a general language for scientific computing, including the concepts of imperative and declarative languages, Python's relationship to other languages, and a brief account of languages widely used in the biosciences; *(ii)* a structured guide to some of the available software packages in computational biology, with an emphasis on Python; and *(iii)* two sample Supplemental Chapters (one basic, one more advanced), along with a brief, practical introduction to the Python interpreter and integrated development environment (IDE) tools such as IDLE.

# Acknowledgments

We thank M. Cline, S. Coupe, S. Ehsan, D. Evans, R. Sood and K. Stanek for critical reading and helpful feedback on the manuscript.

## Figure 1

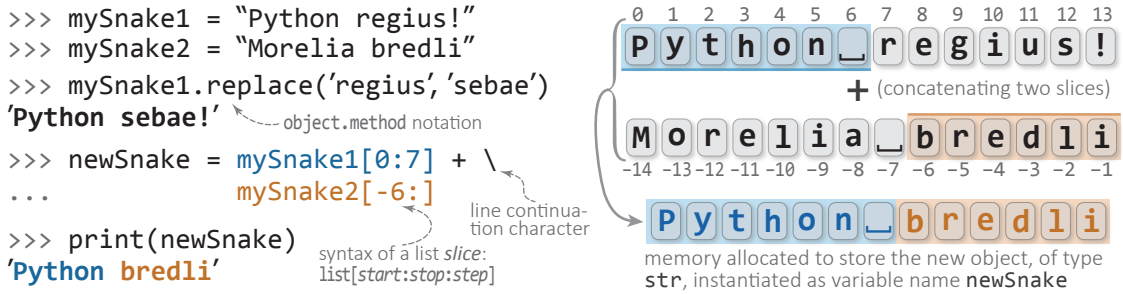

## Figure 2

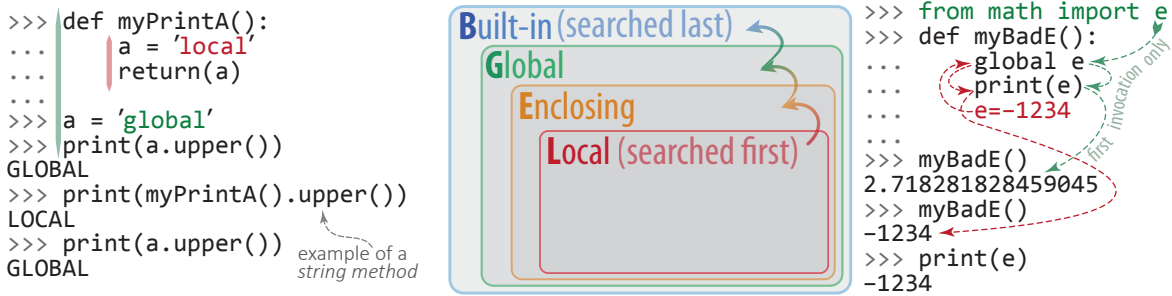

## Figure 3

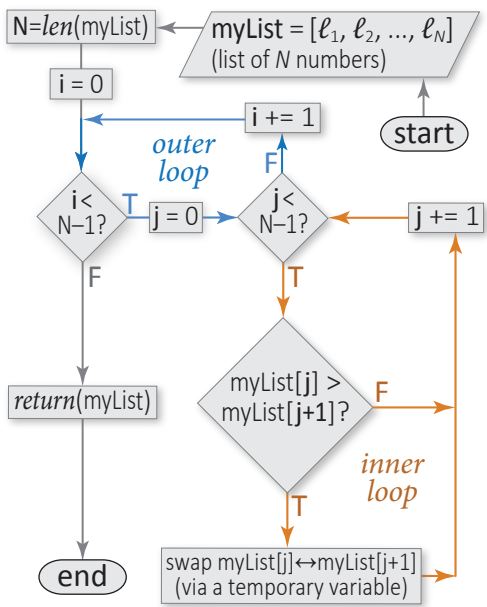

'**Striking Figure**' (logo for http://p4b.muralab.org)

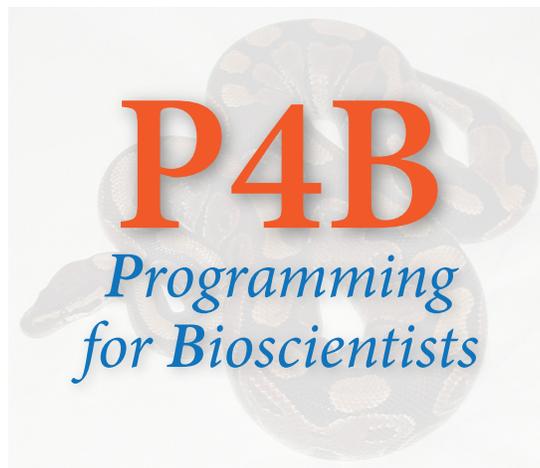

# Supporting Information (S2 Text)

# An Introduction to Programming for Bioscientists: A Python-based Primer

Berk Ekmekci[☯], Charles E. McAnany[☯], Cameron Mura[*]

Department of Chemistry, University of Virginia, Charlottesville, VA 22904-4319 USA;
[*]cmura@muralab.org; [☯]These authors contributed equally to this work.

March 26, 2016

## Contents

# 1 Python and the Broader Scientific Computing Ecosystem

Programming languages can be loosely classified, based on various criteria, into distinct lineages or taxonomies. *Imperative* programming languages require the programmer to specify the detailed steps necessary for the program to run, including explicit instructions for modifying data. In an imperative program, the series of statements that comprise the program alter the runtime's state in a predictable and fairly obvious manner; C and Fortran are examples of popular languages that are often used to program in an imperative manner. The Unix shells (`bash`, `csh`, etc.) are also examples of imperative languages: the user types commands one-by-one and the shell executes them in order as it receives them.

In contrast, *declarative* languages emphasize expressions rather than statements. The programmer specifies not the precise steps necessary to generate an answer, but rather expresses the answer directly in terms of the input to the program. For example, in the Prolog programming language, one defines a set of rules (like a system of algebraic equations) and then asks the interpreter if a certain input could satisfy those rules. The interpreter will either find a valid solution or prove that no solutions exist. Regular expressions specify a set of strings and the regex engine tries to match an input string with a given regex. *Functional* languages are declarative programming languages that consider programs to be functions that can be evaluated. As an example, the parallel pipelines in Python (PaPy) toolkit was written in Python, making extensive use of higher-order functions (e.g., `map`, `reduce`), anonymous or 'lambda' functions (see Chapter 13, Functions III), lazy (non-strict) evaluation as its dataflow model, and other elements of functional programming design [1]. Functional languages discourage, and in some cases prohibit, mutation. While `x=x+1` is valid in most imperative languages (the value of `x` is incremented by one), many functional languages have no easy way to change a variable—the programmer would usually refer to `x+1` when that value was needed.

A number of other programming paradigms can be defined, and many of them overlap. Python is considered a *multi-paradigm* language: it provides many of the tools used in functional programming, including a powerful list comprehension syntax, and it also allows the user to define functions as sequences of statements to be executed (the imperative style).

Regardless of the classification scheme, all programs are characterized by two essential features. As mentioned in the main text, these two characteristics are: (i) **algorithms** or, loosely, the 'programming logic', and (ii) **data structures**, or how data are represented/structured, whether they are mutable, etc. [2] Python treats these two features of a program as inseparable, thereby making it particularly well-suited to the object-oriented programming (OOP) paradigm. Indeed, literally everything is an object in Python.

Python has become an especially popular language in scientific computing largely because (i) its clean syntax and straightforward semantics make it a lucid and readily accessible first language in which to learn/extend/maintain code; (ii) as a language, Python is quite expressive [3, 4], and is inherently amenable to modern programming paradigms such as OOP and functional programming [5]; (iii) Python's widespread popularity has translated into the development of a rich variety of libraries and third-party toolkits that extend the functionality of the core language into every biological domain, including sequence- and structure-based bioinformatics (e.g., BioPython [6]), comparative genomics (e.g., PyCogent [7]), molecular visualization and modelling toolkits (e.g., PyMOL [8], MMTK [9]), 'omics' data-analysis, data processing pipelines and workflow management systems (e.g., [1]), and even parallel programming [10]. Many of these points are further elucidated in [11].

Several languages other than Python have been widely used in the biosciences; see, e.g., [3] for a comparative analysis. The R programming language provides rich functionality for statistical analysis, and has been widely adopted in bioinformatics (e.g., the Bioconductor project [12]). Perl became an early mainstay in bioinformatics programming (e.g., [13, 14]) largely because of its string-processing capabilities (pattern matching, regular expression handling, etc.). The Fortran, C, and C++ languages offer excellent numerical performance with minimal overhead, making them ubiquitous in computationally-intensive tasks such as molecular dynamics (MD) simulation engines; however, these languages require greater care in memory management and other low-level aspects of writing code, versus higher-level languages such as Python or Perl. The D programming language provides performance near that of C, with many convenient language features for high-level programming; however, the resulting language is complex. Though not a suitable tool for numerical computing, Unix shells (`bash`, `csh`, `zsh`, etc. [15]) are often used to link together other standalone programs (shell scripts, Python code, binary executables, etc.) into ad hoc data-processing pipelines.

# 2 A Glimpse of the Bioinformatics Software Landscape

There is a vast array of possible options and starting points for software resources in bioinformatics (and, more generally, computational biology), even if we limit our consideration to software that *(i)* is freely distributed under an open-source license and *(ii)* provides low-level libraries or modular toolkits[†], rather than feature-complete end-products intended for general purpose usage (including by novices). Monolithic, 'all-in-one' software suites typically have many external *dependencies*, and these dependencies generally correspond to low-level libraries; an example from crystallographic computing is the usage of mmdb, Clipper, and various general-purpose graphics libraries (e.g., OpenGL) to achieve the high-level functionality of the popular Coot molecular graphics program [16].

We can only scratch the surface of available software packages, and the subsections that appear below cover but a handful of the programs often encountered in computational biology. The discussion is intentionally biased towards software written in Python, purely for the pedagogical purposes of this primer. Note that the material which appears below is inherently a moving target (and a fast one, at that). It is not uncommon for scientific software projects and databases to be in various states of flux (see, e.g., the editorial in [17])—new packages appear every few weeks, others disappear or become obsolete, and most software codebases undergo extensive modification on the timescale of months. For these reasons, the material in the following subsections strives to point the reader to various lists and *meta-lists* (lists of lists). Such lists are often more stably persistent (e.g., curated on Wikipedia), and they are inherently able to provide more recently updated catalogs of software than can be provided here. Ultimately, a web-search is often the most effective strategy to discover new information, troubleshoot software, ask programming questions, etc.

The remainder of this section is arranged as subsections based on the following major categories: *(i)* Sequence-level bioinformatics, *(ii)* Structural bioinformatics, *(iii)* Phylogenetics and molecular evolution, *(iv)* Omics-scale data-processing, *(v)* Informatics workflow management systems, and *(vi)* The Bio* projects (and some assorted tips). These categories are the common domains of activity in computational biology, both in terms of software development and practical applications. Within each section we offer pointers to online resources that catalog, in an at least somewhat structured way, some of the codes that exist in that application domain; such information often appears as lists and meta-lists.

## 2.1 Sequence-level Bioinformatics

This section's content includes: *(i)* pointers to lists of available sequence analysis software packages that are mostly feature-rich, meaning they can be applied as-is to address a production-grade research task; *(ii)* an example of an educational piece of software ('B.A.B.A.') that covers the dynamic programming algorithm, found in many bioinformatics software packages; and *(iii)* practical advice on locating more detailed information and resources, for Python coding and beyond.

- **Lists of software**: An extensive list of sequence alignment codes is at [18]. A wiki is an ideal format for maintaining oft-changing lists of software, as the information can be readily updated by developers, users, and other members of the scientific community. The wiki content cited above is structured by type of application (pairwise sequence alignment, multiple sequence alignment, sequence motif detection, etc.), and a major subsection is dedicated to software for visualization of alignments [19]. Also, a closely related list is at [20], which supplies some programming tools (typically lower-level than the previous two cited URLs) for statistical computing, of the sort that often factors into sequence alignment methods. For instance, this latter list describes the software 'Orange' as "*an open source machine learning and data mining software (written in Python). It has a visual programming front-end for explorative data analysis and visualization, and can also be used as a Python library.*" Many of the software packages in the above list are open-source, meaning that one can freely access and study the code in order to identify useful chunks of code; these modular units of code can be adapted and re-used for one's own purposes.

---

[†]Software can be described as a *low-level library* or *toolkit* if it provides a generic, modular, and reusable set of functions (e.g., a PDB file parser), independent of specific application domains or highly specific instances of tasks (e.g., splitting a PDB file by chain identifier and writing each chain as a separate file); see also §1 of the main text for more discussion of the terms 'low-level' and 'high-level'.

- **An educational code: B.A.B.A.**: Though written as a Java applet rather than as Python source code, we mention the 'Basic-Algorithms-of-Bioinformatics Applet' (B.A.B.A.; [21]) because of its pedagogical value in learning the dynamic programming algorithm that underlies many sequence-based methods [22]. Given a user-specified input problem (e.g., two sequence strings to align), the B.A.B.A. applet visually builds the dynamic programming matrix. Users can watch the matrix elements be updated as the algorithm progresses, including for such methods as the Needleman-Wunsch algorithm (globally optimal sequence alignment [23]), the Smith-Waterman method (local sequence alignment [24]) at the heart of the BLAST search method, and the Nussinov algorithm (for prediction of RNA secondary structural regions [25]). Using a tool like B.A.B.A., one can learn the dynamic programming algorithm and play with toy-models in preparation for implementing one's own code in Python.

- **Further resources for coding**: For the hands-on activity of actually implementing an algorithm in Python, the most effective and general path to helpful information is a web search, e.g. using Google. By searching the web, one can discover, for instance, valuable and comprehensive discussions of the 'bottom-up' (exhaustive tabulation) and 'top-down' (recursion, memoization) approaches to dynamic programming (see, e.g., [26], [27], and [28]). This same advice holds true for any algorithm or data structure that one is attempting to implement in Python: websites, and online communities of coders, are invaluable resources for both novice and seasoned programmers.

## 2.2 Structural Bioinformatics

Both types of software resources for structural bioinformatics—*(i)* feature-rich suites that can be immediately applied to a research task and *(ii)* lower-level Python libraries that are intended more as modules to incorporate into one's own code—can be discovered and used via similar strategies as mentioned above. Namely, we suggest a combination of *(i)* web-search, *(ii)* consulting lists of software on various wikis and other websites (curated sites are particularly helpful), and *(iii)* inspection of existing code from open-source packages. Some more specific notes follow:

- **Structure alignment/analysis**: As an example of a frequent computational task in structural bioinformatics, consider the comparison of two (or more) 3D structures. There are many available packages for optimal pairwise superimposition of two protein structures; the multiple alignment problem is more difficult (and fewer software solutions exist). Many of the available structural alignment packages are tabulated at [29] and, as of this writing, that web resource offers good coverage of existing packages. To visualize the results of structure alignment calculations, one can find numerous possibilities in such lists as [30] and Table 1 of [31].

- **Python-centric suites**: There are many feature-rich, research-grade software suites available for structural analysis tasks (in many cases, these programs also provide advanced visualization capabilities). Several such programs provide a Python API, or a built-in scripting language or shell that resembles Python's syntax. Examples include the Python-based molecular viewing environment PMV [32], the popular PyMOL molecular graphic program [8], and the macromolecular modelling toolkit (MMTK; [9]). PMV supplies extensive functionality for protein structural analysis, with an emphasis on geometric characteristics (surface curvature, shape properties, etc.). MMTK is "*an open-source program library for molecular simulation applications*", and it provides users with a vast array of Python-based tools. Using MMTK's Python bindings, one can write Python scripts to perform a coarse-grained normal mode calculation for a protein, a simple molecular dynamics (MD) simulation, or molecular surface calculations as part of a broader analysis pipeline.

- **Molecular simulations**: Another type of activity in structural bioinformatics entails molecular modeling and simulation, ranging from simple energy minimization to MD simulations, Monte Carlo sampling, etc. Software packages that are suitable for these purposes are tabulated at [33]. The Bahar lab's 'ProDy' software is an example of a package in this scientific domain that makes substantial use of Python [34]. This "*free and open-source Python package for protein structural dynamics analysis*" is "*designed as a flexible and responsive API suitable for [...] application development*"; this code provides much functionality for principal component analysis and normal mode calculations.

- **Pure Python**: Finally, note that the purely Python-based SciPy toolkit supplies many types of computational geometry utilities that are useful in analyzing macromolecular 3D structures [35]. For instance, the Python module on spatial data structures and algorithms (scipy.spatial [36]) can compute Delaunay triangulations (and, inversely, Voronoi diagrams) and convex hulls of a point-set; this module also supplies data structures, such as $k$D-trees, that are indispensable in the geometric analysis of proteins and other shapes.

## 2.3   Phylogenetics and Molecular Evolution

This section describes software resources for computational phylogenetics, a major goal of which is the calculation of phylogenetic trees that accuratey capture the likely evolutionary history of the entities under consideration (be they protein sequences, genes, entire genomes, etc.).

- Wikipedia's list of phylogenetics packages is quite well-developed [37]. Also, a long-time pioneer of the field, J. Felsenstein, maintains a thoroughly curated list of several hundreds of phylogeny-related software packages at [38]. Notably, the software cataloged at this resource can be listed by methodology (general-purpose packages, codes for maximum likelihood methods, Bayesian inference, comparative analysis of trees, etc.); also, that URL provides a list of pointers to other lists. Many of the phylogeny packages listed on the above web-pages are feature-complete and ready for direct application to a research problem (perhaps in a Python script, depending on the package and its API), while others are libraries that serve as sources of lower-level functionality.

- PyCogent: The comparative genomics toolkit, PyCogent, is an example of a Python-based code in the evolutionary genomics domain. This software package supplies "*a fully integrated and thoroughly tested framework for novel probabilistic analyses of biological sequences, devising workflows, etc.*" [7]. As a concrete example of the benefits of the open-source approach, low-level Python functionality for protein 3D structural analysis was added to PyCogent by third-party developers [39], thereby expanding the scope of this (largely sequence-based) code to include structural approaches to molecular phylogenetics.

- DendroPy: A "*Python library for phylogenetic computing*", DendroPy is a codebase that provides "*classes and functions for the simulation, processing, and manipulation of phylogenetic trees and character matrices*". It also "*supports the reading and writing of phylogenetic data in a range of formats, such as NEXUS, NEWICK, NeXML, Phylip, FASTA, etc.*" [40] DendroPy is described by its authors as being able to "*function as a stand-alone library for phylogenetics, a component of more complex multi-library phyloinformatic pipelines, or as a scripting 'glue' that assembles and* *drives such pipelines.*" This statement perfectly captures the essence of a well-engineered, extensible, open-source scientific software tool, which encourages modularity and code re-use.

- Finally, as an efficient, Python-based approach to developing one's own code in the area of phylogenetics and molecular evolution, the wide-ranging BioPython project (see below) now includes a Bio.Phylo module. This module is described in [41] as supplying "*a unified toolkit for processing, analyzing and visualizing phylogenetic trees in BioPython.*"

## 2.4    Omics-scale Data-processing

The term *omics* refers to the acquisition, analysis, and integration of biological data on a system-wide scale. Such studies have been enabled by the development and application of high-throughput next-generation technologies. Specific sub-fields include, in roughly chronological order of their development, *genomics*, *proteomics*, *metabolomics*, *transcriptomics*, and a panoply of other new omics (*interactomics*, *connectomics*, etc.); the term *NGS* (next-gen sequencing) is closely associated with many of these endeavors. As would be expected, the volume and heterogeneity of data collected on the omics scale present many computing challenges, in terms of both basic algorithms as well as the high-performance software requirements for practical data-processing scenarios. These challenges are largely responsible for spurring the development of many open-source libraries and software packages. Berger et al. [42] recently presented an authoritative review of many computational tools for analyzing omics-scale data, including tables of available software packages sorted by helpful criteria (type of task, type of algorithm). For instance, the review describes potential solutions for data-processing pipelines for transcriptomics data, as obtained from RNA-seq or microarray experiments. An example of an omics-scale software package written chiefly in Python is 'Omics Pipe' [43].

Historically, much of the statistical computing tools that can be used in omics-style bioinformatics has been developed in the R language (see, e.g., the Bioconductor project). A low-level interface between Python and R is available—namely, the RPy package and its recent successor (rpy2) enable the use of R code as a module in Python. As a concrete example of a 'cross-language', integrated omics approach, note that microarray datasets can be processed using established R tools, followed by seamless analysis in Python (via hierarchical clustering) to obtain heat-maps and the corresponding dendrograms [44].

## 2.5    Informatics Workflow Management Systems

A bioinformatics research project is an inherently computationally-intensive pursuit, often entailing complex workflows of data production or aggregation, statistical processing, and analysis. The data-processing logic, which fundamentally consists of chained transformations of data, can be represented as a *workflow*. Several *workflow management systems* (WMS) have been developed in recent years, in biology and beyond, with the goal of providing robust solutions for data processing, analysis, and provenancing[†]. Available libraries and toolkits enable users to create and execute custom data-processing pipelines (in Python), and feature-rich software frameworks also exist. As mentioned in this section (below), some lightweight, production-grade solutions are written entirely in Python.

WMS software is well-suited to the computational and informatics demands that accompany virtually any major data-processing task. A WMS software suite provides the functional components to enable one to create custom data-processing pipelines, and then deploy (*enact*) these pipelines on either local or distributed compute resources (in the cloud, on an e-science grid, etc.). A WMS can be a high-level, feature-rich, domain-independent software suite (e.g., Taverna [45], KNIME [46]), or a lightweight library that exposes modular software components to users (e.g., PaPy [1]). Usage of a WMS is roughly similar in spirit to using, say, a series of low-level Unix scripts as wrappers for data-processing tools; however, compared to the one-off scripting approach, most WMS solutions feature greater flexibility and extensibility, enhanced robustness, and are generally applicable in more than one scientific domain. WMS suites that are often employed in bioinformatics are described, including literature references and software reviews, in an article that reports the creation of a lightweight Python library known as PaPy [1]. PaPy is a purely Python-based tool that enables users to create and execute modular data-processing pipelines, using the functional programming and flow-based programming paradigms. Alongside the PaPy toolkit, other Python-based WMS solutions include Ruffus [47] and Omics Pipe [43].

---

[†]Loosely, *data provenance* involves carefully logging results (and errors, exceptions, etc.), ensuring reproducibility of workflows by automatically recording run-time parameters, and so on.

## 2.6 The Bio* Projects, and Where to Go Next

This final section lists the many available 'Bio*' projects, where the '*' wildcard is a placeholder for a particular programming language; typically, the language at the core of a given Bio* project has seen sufficiently widespread usage in computational biology to warrant the concerted development of general-purpose libraries in that language; notable examples are the the BioPython and BioPerl projects. The table below provides a sampling of these projects, which are cataloged more thoroughly at the Open Bioinformatics Foundation [48]. The table is followed by a few assorted tips and pointers to aid in the discovery of additional resources.

### Table S1: The Bio* projects

| Project | Description (from each respective website) |
|---|---|
| **BioPython** [6] | "Biopython is a set of freely available tools for biological computation written in Python by an international team of developers. It is a distributed collaborative effort to develop Python libraries and applications which address the needs of current and future work in bioinformatics." |
| **BioPerl** [13] | "a community effort to produce Perl code which is useful in biology" |
| **BioJava** [49] | "BioJava is an open-source project dedicated to providing a Java framework for processing biological data. It provides analytical and statistical routines, parsers for common file formats and allows the manipulation of sequences and 3D structures. The goal of the biojava project is to facilitate rapid application development for bioinformatics." |
| **BioSQL** [50] | "BioSQL is a generic relational model covering sequences, features, sequence and feature annotation, a reference taxonomy, and ontologies (or controlled vocabularies)." |
| **BioRuby** [51] | "Open source bioinformatics library for Ruby" |

We conclude by noting the following potentially useful resources:

- Wikipedia's top-level page on bioinformatics software, organized by categories, is at [52]. At a similar level of detail, there exist open-access journals that will be of interest (e.g., *Algorithms for Molecular Biology*); also, *PLOS Computational Biology* publishes new contributions in a dedicated 'Software Collection' [53].

- The bioinformatics.org site [54] includes a list of hundreds of packages for bioinformatics software development; brief annotations for many of these packages are available at a related URL [55]. In addition, ref [56] lists software for Linux, sorted by categories. One can search for 'python' on that page, and will find numerous codes of potential interest.

- Another useful online search strategy is to query PyPI, the Python Package Index [57]. Searching PyPI with terms such as 'bioinformatics' will retrieve numerous potentially useful hits (a beneficial feature of this approach is that the returned codes are likely to be under active, or at least recent, development).

# 3 Supplemental Chapters of Python Code: Two Samples

In addition to the examples of Python code in the main text, a suite of Supplemental Chapters is provided with this work. These freely-available Chapters cover the topics enumerated in Table 1 (main text), and the latest source files are maintained at `http://p4b.muralab.org`. Each Chapter is written as a Python file—i.e., each Chapter is a plaintext `.py` file that can be read, executed, modified, etc. as would any ordinary Python source code. For pedagogical purposes, each Chapter is heavily annotated with explanatory material. These explanations take the form of comments in the Python code (lines that begin with a pound sign, '#'), thereby allowing the Chapters to be both descriptive and executable. The remainder of this section consists of two sample chapters, following a brief subsection that describes some practicalities of interacting with Python in a Unix shell environment. (Python is cross-platform, and various Python interpreters are freely available on the Windows and Apple OS X operating systems too.)

## 3.1 The Python Interpreter, the Unix Shell, and IDEs

Virtually all modern Linux distributions include a recent version of Python. One can begin an interactive Python session by accessing the *interpreter* (see §1 of the main text for more on this term). This, in turn, can be achieved by opening a Unix shell (terminal) and typing the standard command `python`. (On systems with multiple versions of Python installed [not an uncommon scenario], the command `python3` may be necessary—one should experiment with this on one's own Linux system.) As a concrete example, if the first Supplemental Chapter on control flow (the file ch05controlFlowI.py) is present in the current working directory, then its contents can be imported into an interactive Python session by issuing the statement `import ch05controlFlowI` (note the missing '.py' suffix) at the Python prompt. The default Python prompt is indicated by three greater-than signs (`>>>`); for line continuation, the prompt appears as three periods (`...`). There are alternatives to the default Python interpreter, such as the freely-available IPython command shell [58]. In IPython, one can 'load' this file by typing either `import ch05controlFlowI` (as above) or else `run ch05controlFlowI.py` (note the file extension). Another Python interpreter (and source code editor) is IDLE. This integrated development environment (IDE) for Python is fairly straightforward to use, and IDLE is bundled with all standard, freely available distributions of Python. IDLE provides an interactive session and a simple file editor for creating Python source code. The Supplemental Chapters are simple text files that can be loaded into the IDLE editor and executed. Beyond the popular IPython and IDLE, other options for Python IDEs also exist. For instance, many users on the Windows OS use the Anaconda Python distribution, which is packaged with a large number of scientific computing extensions already pre-configured [59].

## 3.2 A Sample Chapter at a Basic Level: Variables [Chap 2]

The following code is Supplemental Chapter `ch02variables.py`, which is an introductory-level treatment of variables.

```
1  """Programming for Biochemists, lessons for Python programming.
2  Copyright (C) 2013, Charles McAnany. This program is free software: you can
3  redistribute it and/or modify it under the terms of the GNU Affero General
4  Public License as published by the Free Software Foundation, either version 3
5  of the License, or (at your option) any later version. This program is
6  distributed in the hope that it will be useful, but WITHOUT ANY WARRANTY;
7  without even the implied warranty of MERCHANTABILITY or FITNESS FOR A
8  PARTICULAR PURPOSE. See the GNU Affero General Public License for more details.
9  You should have received a copy of the GNU Affero General Public License along
10 with this program. If not, see <http://www.gnu.org/licenses/>."""
11 #Chapter 2: Using variables.
12 #We can tell Python to remember the result of a calculation, by storing it
13 #in a variable. The syntax is
```

```python
14  #variableName = calculation
15  #Here, let me make a variable to store seven times seven.
16  def sevSqu():
17      sevenSquared = 7 * 7
18      # I can now use sevenSquared anywhere in this function.
19      print(sevenSquared)
20  
21  #Rules for variables:
22  #Variables must start with a letter or underscore. After that, any combination
23  #of numbers and letters is OK. These are some good variable names:
24  # value1
25  # fooBar
26  # juice_concentration_2
27  #But these are not valid variables.
28  # 1stValue (starts with a number)
29  # let me in (has spaces in it.)
30  # I1lIl1 (technically valid, but I'll kill you if you use this.)
31  
32  ##### NOTA BENE:
33  #Variables in python are case sensitive!
34  # aNumber is a totally different variable than Anumber. This can lead to very
35  #subtle bugs. So:
36  # Use consistent naming schemes. I will almost always capitalize the first
37  # letter of each word in my variables, except the first one. My variables
38  # might be:
39  # aLittleBit
40  # numberOfRottenBananas
41  # counter (only one word, so no capitals.)
42  # (programmers refer to this capitalization scheme as camel casing.)
43  
44  
45  #Some variable use:
46  def onePlusMe():
47      number = 5
48      number = number + 1
49      print(number)
50  # Hm?! Okay, this deserves an explanation.
51  #When python sees this, here's what it will do.
52  # 1. it sees that you're assigning a variable.
53  # 2. it calculates the value you're going to assign. Here, it's number + 1
54  # 3. it sets number to that value.
55  
56  #So remember, = is NOT a question. It's an instruction.
57  #In a mathematics course, if you were told that x=x+1, your initial
58  #response might be along the lines of "no it isn't", and rightly so!
59  #In Python, it means "Calculate number + 1. Number is now that value."
60  
61  #Here's a program to convert temperatures from Celsius to Fahrenheit.
62  def celsToFahr():
63      celsiusTemperature = 30
64      fahrenheitTemperature = 9/5 * celsiusTemperature + 32
65      # Python does order of operations correctly.
66      print("the temperature is ", fahrenheitTemperature, "degrees Fahrenheit.")
67  # Print is a very strange function, in that it can take many arguments
68  # (separated by commas). For now, know that it will glue the arguments
69  # together and print the whole thing.
70  
71  #For some of the programs you'll write, you need to look at the characters
72  #in strings. I'll cover the syntax in greater detail is CH08, collections I,
73  #but for now I'll just give you some useful syntax for strings:
74  #To read the kth character of a string, you say stringName[k-1]
75  #Most of the time, you'll just need the first character, which we can
76  #extract with stringName[0]
77  def stringManip():
78      initialString = "aoeu"
79      print("the string starts with ", initialString[0])
80      #We could get the third character like this:
81      print("The third character is ", initialString[2])
```

```python
82      #It's 2, not 3, because we use k-1 for the number, not k.
83      #(the reason becomes much clearer in CH08)
84
85  #Good? Good.
86
87
88  ####################
89  ##   Exercises   ##
90  ####################
91  # 1. Rewrite the temperature program to convert a Fahrenheit temperature to
92  # a celsius one. What is the celsius temperature when it is 100 F?
93  # Reminder: Celsius = 5/9 (Fahrenheit - 32)
94  def fahrToCels():
95      pass #again, delete the pass, replace this function with your code.
96
97
98
99  #2. Track the value of each of the following variables during this program.
100 #Just fill out the table with the values as they change.
101 #(don't run the code, do it by hand.)
102 def exercise2():
103                     # a | b | c #
104     a = 1           # 1 | ? | ? #
105     b = 1           # 1 | 1 | ? #
106     c = 1           # 1 | 1 | 1 #
107     a = b + c       # 2 | 1 | 1 #
108     b = a + c       # 2 |   |   #
109     c = b + a       #   |   |   #
110     b = c           #   |   | 5 #
111     a = a + b       #   |   |   #
112     c = c * c       # 7 | 5 |   #
113
114 #3. Print the first three characters of the specified string:
115 def printChars():
116     someChars = "aoeuhtns"
```

## 3.3 A Sample Chapter at a More Advanced Level: Classes & Objects, II [Chap 16]

The following code is Supplemental Chapter `ch16ClassesObjectsII.py`, which is a more advanced presentation of classes, objects, and object-oriented programming.

```python
1  """Programming for Biochemists, lessons for Python programming.
2  Copyright (C) 2013, Charles McAnany. This program is free software: you can
3  redistribute it and/or modify it under the terms of the GNU Affero General
4  Public License as published by the Free Software Foundation, either version 3
5  of the License, or (at your option) any later version. This program is
6  distributed in the hope that it will be useful, but WITHOUT ANY WARRANTY;
7  without even the implied warranty of MERCHANTABILITY or FITNESS FOR A
8  PARTICULAR PURPOSE. See the GNU Affero General Public License for more details.
9  You should have received a copy of the GNU Affero General Public License along
10 with this program. If not, see <http://www.gnu.org/licenses/>."""
11 #Chapter 16: Objects and classes: philosophy and iterators.
12 #This is it, the penultimate chapter! You will use everything you've learned
13 #up to this point in this chapter. It should be fun!
14
15 #I'm going to start this chapter by giving you something practical:
16 #an example of a fully-formed class that shows you how classes
17 #are used. A rational number is one of the form a/b, where a and b are
18 #integers. Python does not have built-in support for rationals
19 #and rational arithmetic.
20 class RationalNumber:
21     """A class that implements a rational number and the necessary
22     Arithmetic operations on it."""
23     def __init__(self,numerator, denominator):
24         """Arguments should be numbers or RationalNumbers, and will
25         be the values of this rational number's numerator and denominator."""
```

```python
26            if(isinstance(numerator, RationalNumber)):
27                if(isinstance(denominator, RationalNumber)):
28                    #The constructor was called with RationalNumbers
29                    self._n = numerator._n * denominator._d
30                    self._d = denominator._n * numerator._d
31                else:
32                    #The numerator, but not denominator, is a RationalNumber
33                    self._n = numerator._n
34                    self._d = denominator* numerator._d
35            else:
36                if(isinstance(denominator, RationalNumber)):
37                    #The denominator, but not numerator, is a RationalNumber
38                    self._n = numerator * denominator._d
39                    self._d = denominator._n
40                else:
41                    #Both arguments are plain old numbers
42                    self._n = numerator
43                    self._d = denominator
44            if(self._n != 0):
45                self.reduceFraction()
46            else:
47                self._d = 1
48
49        def reduceFraction(self):
50            gcd = greatestDivisor(self._n, self._d)
51            self._n //= gcd
52            self._d //= gcd
53
54        def add(self, otherNum):
55            """Adds a rational number to this one, using the fact that
56            a/b + c/d = (a*d + c*b)/(b*d)"""
57            return RationalNumber(self._n*otherNum._d+otherNum._n*self._d, self._d*otherNum._d)
58
59        def subtract(self,otherNum):
60            negOther = RationalNumber(-otherNum._n, otherNum._d)
61            return self.add(negOther)
62
63        def mult(self, otherNum):
64            return RationalNumber(self._n * otherNum._n, self._d * otherNum._d)
65
66        def divide(self, otherNum):
67            return RationalNumber(self._n * otherNum._d, self._d * otherNum._n)
68        def __str__(self):
69            return "{0:d}/{1:d}".format(self._n, self._d)
70
71 #I put the code for GCD outside the class - it's not really associated with
72 #rational numbers, so it should be in a different place.
73 def greatestDivisor(a,b):
74     if(b == 0):
75         return a
76     return greatestDivisor(b,a % b)
77
78 def useRational():
79     #a = 1/2
80     a = RationalNumber(1,2)
81     #b = 1/3
82     b = RationalNumber(1,3)
83     #c = a + b
84     c = a.add(b)
85     print(c)
86     #Now to demonstrate that rationals are truly precise...
87     storage = RationalNumber(0,1)
88     floatSum = 0
89     for i in range(1000):
90         storage = storage.add(RationalNumber(1,1000))
91         floatSum += 0.001
92     print(floatSum)
93     print(storage)
```

```
 94        floatZero = floatSum - 1.0
 95        storageZero = storage.subtract(RationalNumber(1,1))
 96        print(floatZero)
 97        print(storageZero)
 98        #The floating point version has some noise that has accumulated during
 99        #the computation. The rational does not have this noise.
100    
101    
102    
103   #Next: Something practical. You know how you can do
104   #for i in range(10):
105   #, right? Well, range is just a class with a few methods defined.
106   #A class is iterable (may be used with a for loop) if it defines the
107   #method __iter__() that returns an object with a method called __next__().
108   #__next__() should return the next value in the sequence or raise
109   #a StopIteration exception.
110    
111   class NewRange():
112       def __init__(self, start, stop):
113           print("NewRange.__init__")
114           self._start = start
115           self._stop = stop
116       def __iter__(self):
117           print("NewRange.__iter__")
118           return RangeIterator(self._start,self._stop)
119    
120   class RangeIterator():
121       def __init__(self,start,stop):
122           print("RangeIterator.__init__")
123           self._currPos = start
124           self._endPos = stop
125       def __next__(self):
126           print("RangeIterator.__next__", end = " ")
127           if self._currPos < self._endPos:
128               self._currPos = self._currPos + 1
129               print(" -> {0:d}".format(self._currPos-1))
130               return self._currPos - 1 #-1 because I already incremented, return
131           else:                         #what the value was, not what it is.
132               print(" -> StopIteration")
133               raise StopIteration
134    
135    
136   #If your class contains a method called __next__(), you can have __iter__
137   #just return self:
138    
139   class SimpleRange:
140       def __init__(self,start,stop):
141           self._currPos = start
142           self._endPos = stop
143       def __next__(self):
144           if self._currPos < self._endPos:
145               self._currPos = self._currPos + 1
146               return self._currPos - 1 #-1 because I already incremented, return
147           else:                         #what the value was, not what it is.
148               raise StopIteration
149       def __iter__(self):
150           return self
151    
152   #When Python comes to a for loop, it first calls __iter__(), then repeatedly
153   #calls __next__() on that iterator until it throws StopIteration.
154   #The advantage is we can just use it like a normal range.
155   def useNewRange():
156       nr = NewRange(0,10)
157       for i in nr:
158           print (i)
159       sr = SimpleRange(0,10)
160       for i in sr:
161           print(i)
```

```python
162
163
164  #Okay, let's get biochemical again. Consider a class that stores DNA:
165  class DNAStore:
166      """Represents a strand of DNA. Accepts new dna as strings or collections
167      of strings. """
168      _bases = "" #Currently empty.
169
170      def __init__(self, bases):
171          """bases is a string or a sequence of strings that will be added to
172          this objects' dna store."""
173          self.add(bases)
174          print("Initialized DNA strand with {0:s}".format(self._bases))
175
176      def add(self, newDNA):
177          """Adds new dna to the end of this strand. Rules for dna are the same
178          as for the initializer."""
179          if isinstance(newDNA, str):
180              for base in newDNA:
181                  self._addLetter(base)
182          elif isinstance(newDNA, (tuple,list)):
183              for thing in newDNA:
184                  self.add(thing) #If it's a tuple or list, split it and add
185                                  #each part of it recursively.
186          else:
187              raise Exception("Invalid DNA.")
188
189      def _addLetter(self, base):
190          if base in "AGTC":
191              self._bases = self._bases + base
192          else:
193              raise Exception("Unknown letter for DNA: {0:s}".format(base))
194
195      def getBases(self):
196          return self._bases
197
198
199  #I'd like to extend this class to allow me to iterate over the codons.
200  class IterableDNA(DNAStore):
201      """An iterable version of a DNA store. Iterates by *codon*, not by
202      *base*."""
203      _bases = "" #Currently empty.
204
205      def __init__(self, bases):
206          """bases is a string or a sequence of strings that will be added to
207          this objects' dna store."""
208          self.add(bases)
209          print("Initialized DNA strand with {0:s}".format(self._bases))
210
211      def add(self, newDNA):
212          """Adds new dna to the end of this strand. Rules for dna are the same
213          as for the initializer."""
214          if isinstance(newDNA, str):
215              for base in newDNA:
216                  self._addLetter(base)
217          elif isinstance(newDNA, (tuple,list)):
218              for thing in newDNA:
219                  self.add(thing) #If it's a tuple or list, split it and add
220                                  #each part of it recursively.
221          else:
222              raise Exception("Invalid DNA.")
223
224      def _addLetter(self, base):
225          if base in "AGTC":
226              self._bases = self._bases + base
227          else:
228              raise Exception("Unknown letter for DNA: {0:s}".format(base))
229
```

```
230     def getBases(self):
231         return self._bases
232
233     def __iter__(self):
234         #Initialize the iteration.
235         self._iterPos = 0
236         return self
237     def __next__(self):
238         start = self._iterPos
239         self._iterPos = start + 3
240         if(len(self._bases) - start < 3):
241             raise StopIteration
242         codon = self._bases[start:start + 3]
243         return codon
244
245 def iterateDNA():
246     idna = IterableDNA("AGTGACTAGTCACTACTAGCATGAGACATGACGAT")
247     for cdn in idna:
248         print(cdn)
249         #The big point here is that the person using your class needn't
250         #think about how the iteration works; it "just works" and is clear
251         #and simple.
252
253 ###################
254 ##   Exercises   ##
255 ###################
256 #1. Add a method to DNAStore that calculates the GC content of its stored
257 #dna.
258
259 #2. Add a method to DNAStore that accepts another DNAStore, and calculates
260 #the Hamming distance between itself and the other strand.
261
262 # 3.Explain the behavior of this function:
263 def rangeMess():
264     def printNest(iterable):
265         for i in iterable:
266             for j in iterable:
267                 print("i = {0}, j = {1}.".format(i,j))
268
269     a = range(0,10)
270     b = NewRange(0,10)
271     c = SimpleRange(0,10)
272     print("built-in range:")
273     printNest(a)
274     print("NewRange:")
275     printNest(b)
276     print("SimpleRange:")
277     printNest(c)
278
279 #4. If you play with IterableDNA, you'll notice it has the behavior of
280 #SimpleRange: You can't nest iteration. Fix it.
281 class BetterIterableDNA:
282     pass
283
284 #5. Implement a deque class. (See CH12, circles() for a brief discussion of
285 #deques.)
286 #It should support these operations
287 #pushLeft(thing) :: appends thing to the left end of the deque.
288 #popLeft()       :: removes the leftmost item from the deque.
289 #peekLeft()      :: returns the leftmost item from the deque.
290 #and their corresponding right-side methods.
291 class Deque:
292     pass
293
294 #I have provided this test method for your use:
295 def testDeque():
296     def checkEqual(a,b):
297         if (a != b):
```

```
298                raise Exception("unequal: {0}, {1}".format(a,b))
299      def checkBroken(op):
300          """Tries to run op (which should be a zero-argument function). If op raises
301      an exception, this catches it and returns gracefully. If op does *not* raise
302      an exception, this raises its own to indicate that the code did not fail."""
303          try:
304              op()
305          except(Exception):
306              print("Error occured as expected.")
307              return
308          raise Exception("Code did not indicate an error.")
309                                          # D1              D2
310      d1 = Deque()                        # <>
311      d1.pushLeft(1)                      # <1>
312      d1.pushRight(2)                     # <1, 2>
313      checkEqual(d1.peekLeft(), 1)        # <1, 2>
314      checkEqual(d1.peekLeft(), 1)        # <1, 2>
315      d1.popLeft()                        # <2>
316      checkEqual(d1.peekLeft(), 2)        # <2>
317      #Can the class support being emptied?
318      d1.popRight()                       # <>
319      #Does the class support strange objects being inserted?
320      d1.pushRight((3,4))                 # <(3,4)>
321      d1.pushLeft("aoeu")                 # <"aoeu", (3,4)>
322      checkEqual(d1.peekRight(), (3,4))#<"aoeu", (3,4)>
323      d2 = Deque()                        # '               <>
324      d2.pushLeft(2)                      # '               <2>
325      #Are multiple objects truly independent?
326      checkEqual(d2.peekRight(), 2)       # '               <2>
327      d1.popLeft()                        # <(3,4)>         <2>
328      d1.popLeft()                        # <>              <2>
329      #Beat up the class a bit...
330      for i in range(10000):
331          d1.pushLeft(i)                  # <10000, 9999, ... 1, 0>
332      for i in range(5000):
333          d1.popRight()                   #<10000, 9999, ... 5001, 5000>
334      checkEqual(d1.peekRight(), 5000)
335
336      d3 = Deque()
337      #Does it indicate a problem if I try to remove or read from an empty deque?
338      checkBroken(lambda:d3.popRight())
339      checkBroken(lambda:d3.peekLeft())
340      #Does the deque still work correctly after I try to manipulate it when
341      #empty?
342      d3.pushLeft(1)
343      checkEqual(d3.peekRight(),1)
344
345 #6. Make your deque class iterable. The iteration should start at the left and
346 #yield all the elements, just like for a list. Iterating should NOT destroy
347 #the deque being used. That is, after I iterate it, I should be able to push
348 #and pop and peek just as before and all the values must be the same. As an
349 #example, the following __next__() would violate this requirement:
350 #def __next__(self):
351 #    if(self._isEmpty()):
352 #        raise StopIteration
353 #    self.popLeft()
354 #    return self.peekLeft()
355 #(Assuming, of course, that self refers to the original deque)
356
357 #(If you implemented your deque well, this should not be hard!) Note: You may
358 #assume that the deque is not modified during the iteration, so, for example,
359 #the behavior of the following code is undefined, and will not be tested:
360 #   for elem in deq:
361 #       deq.popRight() #Undefined behavior: Deque is modified during iteration.
362 #       print(elem)
363 #       elem = elem+1  #Also undefined: I'm trying to modify the elements.
364 #You can assume that the iterator will not be nested; if it works like
365 #SimpleRange, that's okay.
```

```
366
367
368  class IterableDeque(Deque):
369      pass
370
371  #7.
372  #Write a method to stress-test your deque, like the tests above.
373  def testIterableDeque():
374      pass
```

*Programming for Bioscientists*      Supporting Information, S2 Text
20. Wikipedia. List of Statistical Packages; 2016. Available from: http://en.wikipedia.org/wiki/List_of_statistical_packages.

21. Casagrande N. Basic-Algorithms-of-Bioinformatics Applet; 2015. Available from: http://baba.sourceforge.net.

22. Eddy SR. What is Dynamic Programming? Nat Biotechnol. 2004 Jul;22(7):909–910.

23. Needleman SB, Wunsch CD. A general method applicable to the search for similarities in the amino acid sequence of two proteins. J Mol Biol. 1970 Mar;48(3):443–453.

24. Smith TF, Waterman MS. Identification of common molecular subsequences. J Mol Biol. 1981 Mar;147(1):195–197.

25. Nussinov R, Jacobson AB. Fast algorithm for predicting the secondary structure of single-stranded RNA. Proc Natl Acad Sci USA. 1980 Nov;77(11):6309–6313.

26. Guest. Dynamic programming and memoization: bottom-up vs top-down approaches; 2011. Available from: http://stackoverflow.com/questions/6164629/dynamic-programming-and-memoization-bottom-up-vs-top-down-approaches.

27. Voithos. Dynamic programming solution to knapsack problem; 2016. Available from: http://codereview.stackexchange.com/questions/20569/dynamic-programming-solution-to-knapsack-problem.

28. Miller B, Ranum D. Dynamic Programming—Problem Solving with Algorithms and Data Structures; 2014. Available from: http://interactivepython.org/runestone/static/pythonds/Recursion/DynamicProgramming.html.

29. Wikipedia. Structural Alignment Software; 2015. Available from: http://en.wikipedia.org/wiki/Structural_alignment_software.

30. Wikipedia. List of Molecular Graphics Systems; 2016. Available from: http://en.wikipedia.org/wiki/List_of_molecular_graphics_systems.

31. O'Donoghue SI, Goodsell DS, Frangakis AS, Jossinet F, Laskowski RA, Nilges M, et al. Visualization of Macromolecular Structures. Nature Methods. 2010 Mar;7(3s):S42–S55. Available from: http://dx.doi.org/10.1038/nmeth.1427.

32. Sanner M. Python: a programming language for software integration and development. J Mol Graphics Mod. 1999;17:57–61.

33. Wikipedia. List of Software for Molecular Mechanics Modeling; 2016. Available from: http://en.wikipedia.org/wiki/List_of_software_for_molecular_mechanics_modeling.

34. Bakan A, Meireles LM, Bahar I. ProDy: Protein Dynamics Inferred from Theory and Experiments. Bioinformatics. 2011;27(11):1575–1577. Available from: http://bioinformatics.oxfordjournals.org/content/27/11/1575.abstract.

35. Jones E, Oliphant T, Peterson P, et al.. SciPy: Open-source Scientific Tools for Python; 2001–. [Online; accessed 2015-06-30]. Available from: http://www.scipy.org/.

36. Scipy community. Spatial data structures and algorithms; 2016. Available from: http://scipy.github.io/devdocs/tutorial/spatial.html.

37. Wikipedia. List of Phylogenetics Software; 2015. Available from: http://en.wikipedia.org/wiki/List_of_phylogenetics_software.

38. Felsenstein J. Phylogeny Programs; 2014. Available from: http://evolution.genetics.washington.edu/phylip/software.html.

39. Cieślik M, Derewenda ZS, Mura C. Abstractions, Algorithms and Data Structures for Structural Bioinformatics in PyCogent. Journal of Applied Crystallography. 2011 Feb;44(2):424–428. Available from: http://dx.doi.org/10.1107/S0021889811004481.
Ekmekci, McAnany, Mura      18 of 19